\documentclass[12pt]{iopart}
\usepackage[english]{babel}

\usepackage{braket}
\usepackage{breqn}
\usepackage{iopams}
\usepackage{graphicx}

\usepackage{color}
\definecolor{darkblue}{rgb}{0,0,0.6}
\definecolor{darkred}{rgb}{0.6,0,0}
\definecolor{darkgrey}{rgb}{0.6,0.6,0.6}
\usepackage[colorlinks=true,urlcolor=darkblue,citecolor=darkblue,linkcolor=darkred,hyperfootnotes=false]{hyperref}

\usepackage{bm}

\newcommand{\beq}{\begin{equation}}
\newcommand{\eeq}{\end{equation}}

\newcommand{\ee}{\mathrm{e}}

\newcommand{\hata}{\hat{a}}
\newcommand{\hatb}{\hat{b}}

\newcommand{\Ac}{\mathcal{A}}
\newcommand{\Bc}{\mathcal{B}}

\newcommand{\Ai}{{\rm Ai}}
\newcommand{\Bi}{{\rm Bi}}

\newcommand{\eps}{\epsilon}

\begin{document}
\title%[short title]
{Dynamical phase coexistence in the Fredrickson-Andersen model}
\author{Robert L Jack$^{1,2}$, Takahiro Nemoto$^{3,4}$, Vivien Lecomte$^5$}
%\author{authors}
\address{$^1$ DAMTP, Centre for Mathematical Sciences, Wilberforce Road, Cambridge CB3 0WA, United Kingdom}
\address{$^2$ Department of Chemistry, Lensfield Road, Cambridge CB2 1EW, United Kingdom}
\address{$^3$ Philippe Meyer Institute for Theoretical Physics, Physics Department, \'Ecole Normale Sup\'erieure \& PSL Research University, 24 rue Lhomond, 75231 Paris Cedex 05, France}
\address{$^4$ Mathematical Modelling of Infectious Diseases Unit, Institut Pasteur, 25-28 Rue du Docteur Roux, 75015 Paris, France.}
\address{$^5$ LIPhy, Universit\'e Grenoble Alpes \& CNRS, F-38042 Grenoble, France}
%\date{2018}

\begin{abstract}
We analyse a first-order dynamical phase transition that takes place in the Fredrickson--Andersen (FA) model.
We construct a two-dimensional spin system whose thermodynamic properties reproduce the dynamical large deviations 
of the FA model and we analyse this system numerically, comparing our results with finite-size scaling theory.  This allows us to rationalise recent results for the FA model, including the exponential divergence of its susceptibility at phase coexistence.  We also discuss a simple interfacial model that reproduces quantitatively the behaviour of the FA model at coexistence.    
\end{abstract}

%\maketitle

%\tableofcontents

\section{Introduction}

Large deviation theory is an increasingly useful tool for understanding dynamical fluctuations in statistical mechanics~\cite{denH-book,Derrida2007,Lecomte2007,Garrahan2009,Touchette2009,Bertini2015}.  
For example, by analysis of probability distributions of time-averaged quantities, one gains understanding of entropy production  and dissipation~\cite{Lebowitz1999,Gingrich2016,Nemoto2019}, transport phenomena~\cite{Bodineau2004,Derrida2007,Baule2008,Hurtado2014,Gao2019}, and metastability in  glassy systems~\cite{Garrahan2007,Hedges2009,Jack2010,Elmatad2010}.
The tails of these probability distributions can have characteristic forms, which are related to a particular type of dynamical phase transition~\cite{Bodineau2004,Garrahan2007,Hedges2009}.  These are analogous to thermodynamic phase transitions, in that one analyses ensembles of dynamical trajectories in a manner that is exactly analogous to the classical analysis of thermodynamic ensembles.  While the principles of thermodynamics are restricted to static and quasistatic phenomena, applications of large deviation theory to trajectories of these systems can be viewed as an extension of the thermodynamic formalism to the analysis of dynamical phenomena~\cite{ruelle1978thermodynamic,Merolle2005,Lecomte2007,Garrahan2009}.

The analogy between static and dynamical phase transitions means that methods from thermodynamics can often be generalised, in order to study dynamical transitions.  For example, the dynamical analogue of the free energy can be obtained as the solution to a eigenvalue problem~\cite{Derrida2007,Touchette2009,Chetrite2015}, just as the thermodynamic free energy can be obtained as the largest eigenvalue of a transfer matrix.  The numerical method of transition path sampling~\cite{Bolhuis2002} generalises Monte Carlo sampling techniques from static configurations to dynamical trajectories~\cite{Merolle2005,Jack2006jcp}.  However, there are also methods that harness directly the dynamical aspects of trajectory ensembles, including cloning (population dynamics) methods for numerics~\cite{Giardina2006,Lecomte2007b}, and control-theoretic approaches that aim to describe dynamical rare events by constructing controlled processes where these events become typical~\cite{Jack2010,Jack2015b,Chetrite2015var}. 

In a recent paper~\cite{Nemoto2017}, we analysed the finite-size scaling of a first-order dynamical phase transition in the Fredrickson--Andersen (FA) model~\cite{Fredrickson1984}.  This simple model describes glassy systems~\cite{Garrahan2002,Chandler2010} and has been  studied extensively in that context.  Its dynamical phase transition is also amenable to analytic and numerical studies~\cite{Garrahan2007,Garrahan2009,Elmatad2010,Bodineau2012cmp,Bodineau2012jsp,Nemoto2017,banuls2019}.  In~\cite{Nemoto2017}, we used a cloning method to demonstrate an exponential divergence of the susceptibility with system size, and we explained this behaviour analytically.
This exponential divergence stands in contrast to the usual finite-size scaling behaviour at \emph{thermodynamic} first-order transitions, where one observes a power-law divergence with system size. 
The different scaling behaviour arises from the way in which the thermodynamic limit is taken: when considering the distribution of a time-averaged quantity, the large-time limit is taken before the large system-size limit, requiring the analysis of two distinct scaling variables.
In this work, we analyse in more detail the connection between finite-size scaling analyses for thermodynamic and dynamical phase transitions.

Our results consist of two main parts.  First, we define a two-dimensional ($2d$) spin model whose behaviour can be related directly to trajectories of the one-dimensional FA model.  This spin model has a thermodynamic phase transition that reproduces all aspects of the FA dynamical phase transition.  We present numerical results that illustrate this fact.  In particular, we show that by considering systems on lattices with different aspect ratios, we can capture the differences in finite-size scaling behaviour  between  the usual thermodynamic setting (aspect ratio close to unity) and the dynamical one (very large aspect ratio).  This observation is related to a theoretical analysis of thermodynamic finite-size scaling by Privman and Fisher~\cite{Privman1983} and Borgs and Koteck\'y~\cite{borgs_rigorous_1990,borgs_finite-size_1992}.  In the second part of this paper, we revisit the analytical results of~\cite{Nemoto2017}, which concern an interfacial model that captures quantitatively the finite-size scaling behaviour of the FA model.  We show how some of the results can be simplified by taking a suitable limit, and we discuss the physical interpretation of this model in more detail.
We explain that our results for finite-size scaling behaviour are generic for first-order dynamical transitions.

The paper is structured as follows: Sec.~\ref{sec:fa} reviews the behaviour  of the FA model and its phase transition.  Sec.~\ref{sec:spin-model} introduces the  $2d$ spin model and Sec.~\ref{sec:numerics} has the results for this model.  Sec.~\ref{sec:interfacial} analyses the interfacial model.  Our conclusions are summarised in Sec.~\ref{sec:conc}.

%The thermodynamic analogue of this analysis is to consider systems with anisotropic shape, whose spatial extent in one direction is much larger than all other directions.
%
%This paper does two main things.
%
% (i) We define a $2d$ spin model whose configurations correspond to trajectories of a discrete-time FA model  For this model we analyse finite-size scaling of the relevant phase transition and we discuss the connection to finite-size scaling for dynamical phase transitions.  (ii) We revisit the analysis of the interfacial model of~\cite{Nemoto2017}, and we discuss its physical interpretation and its connection to numerical results for the FA model and the $2d$ spin model.

\section{Large deviations in the FA model}
\label{sec:fa}

This Section recalls the definition of the FA model~\cite{Fredrickson1984} and summarises some previous results for its dynamical phase transition~\cite{Garrahan2007,Garrahan2009}.

\subsection{Model}

The (one-dimensional) FA model is a Markov chain in continuous time.  It consists of $L$ spins on a $1d$ lattice with periodic boundaries, the $i$th spin is $n_i\in\{0,1\}$.  Let $\bm{n}=(n_1,n_2,\dots,n_L)$.  Spin $i$ can flip (that is, change its state) only if a constraint function $C_i(\bm{n})$ is non-zero.  In this work we take $C_i(\bm{n})=n_{i+1}+n_{i-1}-n_{i+1}n_{i-1}$ so that $C_i\in\{0,1\}$.   We also introduce a numerical parameter $c\in(0,1)$.  Then, spins with $n_i=0$ flip to $n_i=1$ with rate $c\, C_i(\bm{n})$; spins with $n_i=1$ flip to $n_i=0$ with rate $(1-c) C_i(\bm{n})$.   
We note in passing that some other studies take instead $C_i=n_{i-1}+n_{i+1}$ which leads to very similar behaviour.  Our choice in this work is convenient because it simplifies the spin model defined in Sec.~\ref{sec:spin2d-def} below.
Note also that no spin flips can ever cause the system to enter or leave the state $\bm{0}=(0,0,\dots,0)$.  Hence, in order to ensure that the dynamics is irreducible, we define its configuration space as the set of all spin configurations except for $\bm{0}$.

The FA model respects detailed balance with respect to a steady-state (equilibrium) probability distribution where all spins are independent, and $\langle n_i \rangle_0=c$. That is, in  the equilibrium state one has
\beq
p(\bm{n}_t) \propto \exp\left( -(h/T) \sum_i n_{i,t} \right),
\label{equ:pn}
\eeq
where $(h/T) = \log[(1-c)/c]$; we identify $T$ as the temperature.
[To be precise, (\ref{equ:pn}) is applicable only if $\bm{n}_t\neq\bm{0}$.]

\subsection{Large deviations}
\label{sec:ldt}

We briefly describe the large-deviation formalism for the dynamical activity.  For full details see~\cite{Garrahan2009}, general discussions of large deviation theory can be found in~\cite{denH-book,Touchette2009}.
Consider a dynamical trajectory where the time $t$ runs from $0$ to $\tau$.  Let $K(\tau)$ be the total number of spin flip events that occur during this time interval.  The average $\langle K(\tau)\rangle_0 = \tau \overline{k} L $ with $\overline{k} = 2c^2(2-c)(1-c)$.  Since the model is a finite irreducible Markov chain,
 the probability distribution for $K(\tau)$ obeys an large-deviation principle (LDP)~\cite{denH-book,donsker_asymptotic_1975,Lecomte2007,Garrahan2009,Touchette2009}:
\beq
\mathrm{Prob}[K(\tau) \approx \tau k L] \asymp \exp[ -\tau I_L(k) ],
\label{equ:ldp}
\eeq
where $I_L$ is the rate function, which is analytic and strictly convex.  Note that the speed of this LDP is $\tau$ and that the system size $L$ is fixed as $\tau\to\infty$.

Large deviation theory describes rare events where $K(\tau)$ takes a value far from its mean.  
To analyse these rare events it is convenient to introduce a biasing field $s$ and to define the scaled cumulant generating function (SCGF)
\beq
G_{L,\tau}(s) = \frac{1}{\tau} \log \big\langle \ee^{-sK(\tau)} \big\rangle_0 \; .
\label{equ:GLt}
\eeq
The angle brackets with subscript $0$ indicate a steady-state (equilibrium) average over trajectories running between time $0$ and $\tau$. 
Also define 
\beq
G_L(s) = \lim_{\tau\to\infty} G_{L,\tau}(s) \; .
\label{equ:GL}
\eeq
Then the rate function can be obtained as %\emph{(check signs etc)}
$I_L(k) = \sup_s [  - skL - G_L(s) ]$.  Also, the rare events of interest can be analysed by considering biased ensembles of trajectories in which the average of any (trajectory-dependent) observable $O$ is
\beq
\langle O \rangle_s = \frac{ \big\langle O \ee^{-sK} \big\rangle_0 }{ \big\langle \ee^{-sK} \big\rangle_0 } .
\label{equ:Os}
\eeq
Results for equivalence of ensembles~\cite{Lecomte2007,Chetrite2013} show that for a large class of observables $O$, the biased average $\langle O\rangle_s$ converges (as $\tau\to\infty$) to a conditional average of $O$ in the original (unbiased) process, where $K(\tau)$ is constrained to take a particular value.  This constraint is $K(\tau)=\tau L \cdot k_L(s)$ with 
$
%k_L(s)=-G_L'(s)L  .
k_L(s)=-G_L'(s)/L  .
$

Let $p_{L,\tau}(K)$ be the  probability to observe exactly $K$ spin flips and let
$
\rho_{L,\tau}(k) = \tau L \cdot p_{L,\tau}(\tau L k)
$
be an associated probability density for $k=K/(L\tau)$.  The rate function that appears in (\ref{equ:ldp}) may be obtained as
\beq
I_L(k) = \lim_{\tau\to\infty} \frac{-1}{\tau} \log \rho_{L,\tau}(k) \; .
\label{equ:IL}
\eeq
%under the assumption that the limit exists (which is the case in the FA model).

\begin{figure}
\hspace{25mm}\includegraphics[width=125mm]{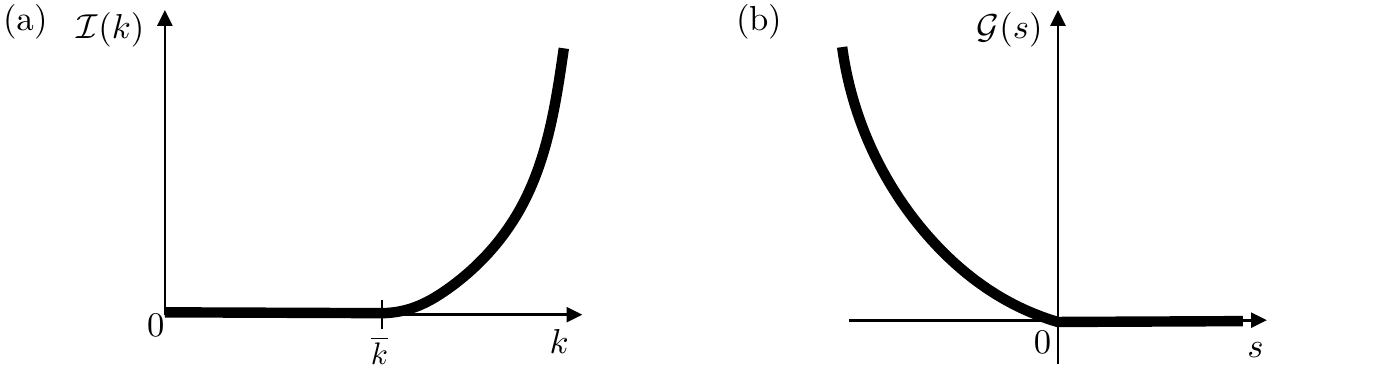}
\caption{Sketches of ${\cal I}(k)$ and ${\cal G}(s)$ for the FA model, based on the theory of~\cite{Garrahan2007,Garrahan2009}.  The first derivative of ${\cal I}$ is continuous at $k=\overline{k}$ but the second derivative is discontinuous there. (Note ${\cal I}'(k)=0$ for $k\leq\overline{k}$.)  
 The first derivative of ${\cal G}$ is discontinuous at $s=0$ (with ${\cal G}'(s)=0$ for $s>0$).}
\label{fig:IG-bulk-sketch}
\end{figure}

\subsection{Dynamical phase transition}

Since the FA model is an irreducible finite-state Markov chain, the Perron--Frobenius theorem establishes that $I_L(k)$ and $G_L(s)$ are analytic and (strictly) convex~\cite{Lecomte2007,Touchette2009,Chetrite2015}.  To reveal the dynamical phase transition that occurs in the model, one defines a bulk free energy density and a scaled rate function
\beq
{\cal G}(s) = \lim_{L\to\infty} \frac{1}{L} G_L(s), \qquad {\cal I}(k) = \lim_{L\to\infty} \frac{1}{L} I_L(k) \; .
\label{equ:GG-II}
\eeq
The form of these limiting functions ${\cal G},{\cal I}$ is not constrained by the Perron--Frobenius theorem, see~\cite{Jack2019-colloq} for a discussion.  In the FA model, these functions have singularities, as sketched in Fig.~\ref{fig:IG-bulk-sketch}.  Note in particular that ${\cal I}(k)=0$ for $0<k\leq\overline{k}$ and ${\cal G}(s)=0$ for all $s\geq0$~\cite{Garrahan2007,Garrahan2009,Bodineau2012cmp}.  In the analogy with equilibrium statistical mechanics, this corresponds to a first-order phase transition.  
%In the following sections we discuss this analogy and we show how this singularity (which appears in a limit of large $L,\tau$) has consequences for rare-event probabilities when $L,\tau$ are both finite.

\section{From the $1d$ FA model to a $2d$ spin system}
\label{sec:spin-model}

In this Section we formulate a discrete-time FA (dFA) model with synchronous spin updates.  The time between updates is $\delta t$ and taking $\delta t\to0$ recovers the (original) FA model.  We also explain how trajectories of this model can be mapped onto configurations of a $2d$ spin system.  Hence dynamical large deviations of the dFA and FA models can be analysed by computing thermodynamic properties of this $2d$ spin system.

\subsection{Discrete-time FA model}

The configuration of the discrete-time FA model at time $t$ is $\bm{n}_t$.  The configuration at time $t+\delta t$ is generated by the following (stochastic) procedure.  We first construct a set of random variables $\bm{m}_t=(m_{1,t},m_{2,t},\dots,m_{L,t})$ with $m_{i,t}\in\{0,1\}$.  The idea is that if $m_i=0$ then spin $n_i$ cannot change its state between times $t$ and $t+\delta t$.  The $m$-variables are chosen from a distribution (independent of $t$ and $\bm{n}$):
\beq
p(\bm{m}) = \frac{1}{z(\gamma)}  \exp\left(-\gamma \sum_i m_i\right)  \prod_i \tilde{C}_i(m_i,m_{i+1})
\eeq
where $\gamma$ is a parameter and $z(\gamma)$ is a normalisation constant (see below); also $\tilde{C}(m_i,m_{i+1})=1-m_{i}m_{i+1}$, which enforces that adjacent sites cannot both have $m=1$.

Given the vector $\bm{m}_t$, the spin variables $n_{i,t+\delta t}$ are then generated independently, by the following rule: 
\begin{eqnarray}
\fl \qquad \hbox{If } C_i(\bm{n}_t) =0 \hbox{ or }  m_i=0 \hbox{ then } & n_{i,t+\delta t} = n_{i,t} 
 \nonumber \\
 \qquad\qquad\hbox{otherwise} &  n_{i,t+\delta t} = 1,0 & \hbox{ with probs } c,(1-c) \hbox{ respectively} 
\end{eqnarray}
The dependence on $m_i$ together with the constraint $\tilde C$ means that if spin $n_i$ flips on a given time step then neither spin $n_{i-1}$ nor spin $n_{i+1}$ can flip.  Hence, if spin $i$  flips then $C_i(\bm{n}_t)=1=C_i(\bm{n}_{t+\delta t})$.  This last property can be used to show that the update rule respects detailed balance with respect to the same steady state distribution as the FA model.  
%If $m_{i,t}=0$ or the FA model constraint $C_i(\bm{n}_t)=0$ then spin $i$ does not flip so $n_{i,t+\delta t}= n_{i,t}$.  If $m_{i,t}=1$ and $C_i(\bm{n}_t)=1$ then spin $i$ may flip: we set $n_{i,t+\delta t}=1$ with probability $c$ and $n_{i,t+\delta t}=0$ with probability $(1-c)$, independent of $n_{it}$.

Positive values of the parameter $\gamma$  increase the probability that $m_i=0$ and reduce the number of sites at which the $n$-spins can flip.  It is easily seen that $z(\gamma)$ is the partition function for a gas of $1d$ hard rods with length 2 and chemical potential $-\gamma T$, on a lattice of size $L$.
For large $\gamma$, the typical number of spins that are able to flip in a single update is small, and $z$ is close to the partition function for an ideal gas.  If one takes
%Returning to the update rule for the discrete-time FA model, 
%It is simple to show that this discrete-time update rule for $\bm{n}$ obeys detailed balance with respect to the steady-state distribution of the FA model.  To recover the continuous-time FA model, it is necessary to take $\delta t\to0$, and to enforce that the number of spins that flip in any time step is proportional to $\delta t$.   
%This can be achieved by fixing $\gamma$ such that $\langle m_i\rangle=\delta t$.  For small $\delta t$ this requires 
\beq
\gamma = - \log \frac{\delta t}{1-\delta t} \; 
\label{equ:h-dt}
\eeq
and $\delta t\to0$ then the dFA model reproduces the behaviour of the FA model.

\subsection{Corresponding spin model}
\label{sec:spin2d-def}

\begin{figure}
\hspace{2cm}\includegraphics[width=125mm]{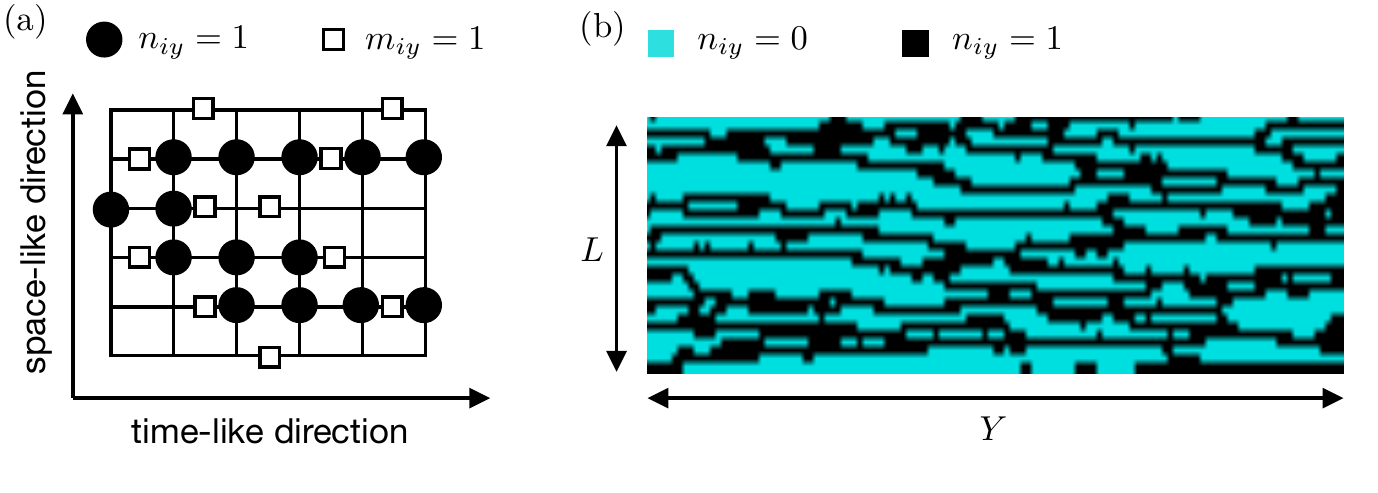}
\caption{Illustration of the $2d$ spin model.  (a) The model is defined on a square lattice.  It is anisotropic, with a space-like direction corresponding to the $1d$ lattice of the analogous FA model and a time-like direction that corresponds to the time axis for the FA model.  
The vertices (sites) of the square lattice are occupied by $n$-variables, and sites with $n_{iy}=1$ are indicated with filled circles.  All other vertices have $n_{iy}=0$.  The horizontal (time-like) bonds are occupied by $m$-variables: 
bonds with $m_{iy}=1$ are indicated by open squares. 
(b) A representative configuration of this model obtained by Monte Carlo sampling with $c=0.38$, $\gamma=0$, and $(L,Y)=(32,128)$.  Only the $n$-variables are shown.  This resembles a typical trajectory of the continuous-time FA model~\cite{Garrahan2002}.
}
\label{fig:spin-model}
\end{figure}

We now define the $2d$ spin model whose configurations correspond to trajectories of the  dFA model.  For a trajectory of length $\tau$, let $Y=\tau/\delta t$ be the number of time steps in the discrete-time FA model. The $2d$ model is defined on a decorated lattice, illustrated in Fig.~\ref{fig:spin-model}(a).  The lattice contains $(L\times Y)$ primary sites, which are associated with variables $n_{i,y}\in\{0,1\}$ with $i=1,2,\dots L$ and $y=1,2,\dots Y$.  The index $y$ indicates the position along the \emph{time-like} axis (horizontal in Fig.~\ref{fig:spin-model}) and the index $i$ is the position on the space-like axis.  The $m$-variables of the  dFA model are defined on the time-like bonds of the lattice.  They are $m_{i,y}\in\{0,1\}$ with $i=1,2,\dots L$ and $y=1,2,\dots, (Y-1)$.

The configurations of this model are in exact correspondence with trajectories of the discrete-time FA model, as long as one enforces the following constraints: the FA configuration $\bm{0}$ is not allowed so $\sum_i n_{i,y}>0$ for all $y$; the $m$-variables on spatially adjacent sites cannot both be in state $1$ so $m_{i,y} m_{i+1,y}=0$ for all $y$; an FA spin can only flip if $m_{i,t}=1$ and $C_i(\bm{n}_{t})=1$, so $\delta(n_{i,y}-n_{i,y+1}) + m_{it} (n_{i-1,t} +n_{i+1,t}) > 0$ where $\delta(x)=1$ if $x=0$ and $\delta(x)=0$ otherwise.  These constraints define the configuration space of the spin model.

We now define an energy function so that the Boltzmann--Gibbs distribution of the $2d$ model at temperature $T=1$ corresponds to the steady-state trajectory measure for the discrete-time FA model.  
This energy is 
\beq
E_0 = -\log \rho_0(\bm{n}_1) - \sum_{y=1}^{Y-1} \log P(\bm{n}_{y+1}|\bm{n}_y) \;,
\eeq
 where $\rho_0(\bm{n})$ is the steady-state probability of configuration $\bm{n}$ in the FA model, and $P(\bm{n}|\bm{n}')$ is the transition probability for the discrete-time dynamics.  The result is
%This energy is 
%leads to a  distribution over spin configurations that corresponds to the .  This is
%\emph{(define $\beta=-\log\frac{c}{1-c}$? this needs a check in any case, may want to add a couple more steps in the derivation.)}
\begin{eqnarray}
\fl \qquad E_0 = \sum_{i=1}^L \sum_{y=1}^{Y-1} \left\{ \gamma m_{it} + C_i(\bm{n}_y) m_{i,y} \left[ \log\frac{1}{1-c} - \frac12 (n_{i,y+1}+n_{i,y}) \log\frac{c}{1-c} \right] \right\}
\nonumber \\ - \frac12 \sum_{i=1}^L (n_{i,1} + n_{i,Y}) \log\frac{c}{1-c}  \; .
\label{equ:E0}
\end{eqnarray}
In deriving this, it is useful to note that the constraints on the configuration space of the model imply that 
$(n_{i,y+1}-n_{iy}) = C_i(\bm{n}_y) m_{i,y}  (n_{i,y+1}-n_{iy})$;  on summing this equation over $y$ the left hand side reduces to $n_{i,Y}-n_{i,1}$.
One sees from (\ref{equ:E0}) that the energy is symmetric under 
 reversal of the time-like direction $(n_{i,y},m_{i,y})\to (n_{i,Y+1-y},m_{i,Y-y})$.  This symmetry is the same as time-reversal symmetry of the steady state of the dFA model, which follows from its detailed balance property.

Correspondence between the continuous-time FA model and this spin model requires that $\gamma$ is large, by (\ref{equ:h-dt}).  This means that sites with $m_i=1$ are rare and also that for a given trajectory length $\tau$, it is necessary to use a very large lattice (because $Y$ is large).  However,  numerical simulations are inefficient in this case.  All qualitative features of this model depend weakly on $\gamma$, so we focus in the following on $\gamma=0$.

Fig.~\ref{fig:spin-model}(b) shows a representative configuration of the spin model with $c=0.38$ and $\gamma=0$, generated by Monte Carlo sampling (see \ref{app:mc}).  Since the mapping between the discrete-time FA model and the spin model is exact, this is also a representative trajectory of the dFA model.  (A dynamical simulation of the dFA model would be a much simpler way to generate a similar trajectory, but this figure illustrates proof of principle.)  The trajectory shown in Fig.~\ref{fig:spin-model}(b) is also consistent with the behaviour of the FA model, see for example~\cite{Garrahan2002}.

\subsection{Free energy and large deviation formalism}

The biased trajectory ensembles defined in (\ref{equ:Os}) can be analysed by considering the $2d$ spin model with a modified energy function.  For any given $2d$ spin configuration, the number of spin flips in the corresponding dFA model trajectory is
\beq
K=\sum_{i=1}^L  \sum_{y=1}^{Y-1}  [ n_{iy} (1-n_{i,y+1}) +  (1-n_{i,y}) n_{i,y+1} ] \; .
\label{equ:K-spin}
\eeq 
  Hence the biased trajectory ensemble for the discrete-time FA model corresponds to a Boltzmann--Gibbs distribution of the $2d$ spin model, where the energy $E_0$ is replaced by 
\begin{eqnarray}
E_s &=& E_0 + sK 
\nonumber\\ 
&=& E_0 + s \sum_{i=1}^L \sum_{y=1}^{Y-1} \left[ n_{iy} +  n_{i,{y+1}} - 2 n_{iy} n_{i,y+1} \right] \; .
\label{equ:Es}
\end{eqnarray}
For $s>0$, this represents an additional ferromagnetic coupling for the spins, along the time-like bonds.
The free energy difference (per site) between systems with energy $E_s$ and $E_0$ is a scaled cumulant generating function:
%
%
%We also define analogues of the quantities introduced in Sec.~\ref{sec:ldt}.  To distinguish dynamical properties of the FA model from equilibrium properties of the $2d$ spin model, we use hats (carets) for the equilibrium case, so
\beq
\hat{G}_{L,Y}(s) = \frac{1}{LY} \log \left\langle \ee^{-sK} \right\rangle_0 \; ,
\eeq
where the angle brackets with subscript zero now indicate a Boltzmann average with energy $E_0$.  
This free energy $\hat{G}$ is analogous to (\ref{equ:GLt}), but note the different normalisation: in the dynamical case we normalised by the trajectory length $\tau$ (analogous to $Y$ in this case);  here we also normalised by $L$, to obtain a free energy per site. 
We use carets (hats) to distinguish functions for the $2d$ spin model from their counterparts in the original FA model.   

Also define
$\hat\rho_{L,Y}(k)$ as the probability density for $k=K/(LY)$ [analogous to $\rho_{L,\tau}(k)$ in Sec.~\ref{sec:ldt}].
The analogue of (\ref{equ:GG-II}) is 
\beq
\fl \qquad
\hat{\cal G}(s) = \lim_{L\to\infty} \lim_{Y\to\infty}  \hat{G}_{L,Y}(s) , 
\qquad
\hat{\cal I}(k) = \lim_{L\to\infty} \lim_{Y\to\infty}  \frac{-1}{LY} \log \hat\rho_{L,Y}(k)
\label{equ:hat-GG-II}
\eeq
We identify $\hat{\cal G}$ as the bulk free energy density.
Note that there are no long-ranged terms in the energy function and the spin model has no other pathological features so we expect on general thermodynamic grounds that the order of limits should be irrelevant in these definitions, and that $\hat{\cal G}(s)$ and $\hat{\cal I}(k)$ should be related by Legendre transformation.  We have not proven these results rigorously but all our results (and previous results for the FA model) are consistent with this hypothesis.

\subsection{Periodic boundary conditions}

In numerical studies of the dynamical phase transition that occurs in the FA model and other glassy models~\cite{Garrahan2007,Garrahan2009,Hedges2009}, finite-size scaling analyses are affected by the initial conditions in the trajectory averages of (\ref{equ:GLt}, \ref{equ:Os}).  For the $2d$ spin model considered here, the equilibrium ensemble of trajectories of the dFA model corresponds to taking free boundary conditions at $y=1$ and $y=Y$, as in (\ref{equ:E0}, \ref{equ:K-spin}).  

However, for the spin model, it is much more natural to consider a fully periodic system, instead of free boundary conditions.  This also helps to reduce finite-size effects.  To achieve this, we introduce an extra column of $m$-variables to the picture in Fig.~\ref{fig:spin-model}(a), which  are $\bm{m}_Y=(m_{1,Y},m_{2,Y},\dots,m_{L,Y})$.  We also periodize (\ref{equ:E0}, \ref{equ:K-spin}) by adding extra terms that couple the spins $\bm{n}_1,\bm{n}_Y,\bm{m}_Y$.   In terms of the dFA model, this means that we consider the ensemble of trajectories that are periodic in time, see also~\cite{Biroli2001}.  Thermodynamic arguments indicate that the bulk free energy density should be $\hat{\cal G}$ independent of boundary conditions.  This result can be confirmed in  this case because
\beq
\hat{G}_L(s) = \lim_{Y\to\infty} \hat{G}_{L,Y}(s)
\eeq
can be obtained as the largest eigenvalue of a particular matrix~\cite{Lecomte2007,Garrahan2009}  (the tilted generator~\cite{Chetrite2015}), and this result is independent of whether one uses periodic or free boundary conditions at $y=1,Y$.

All numerical results for the $2d$ spin model use periodic boundary conditions, except where explicitly stated.   Details of the simulation methods are given in~\ref{app:mc}, including some results with open boundaries.  Note that some care is required with simulation of this spin model, because Monte Carlo moves that flip single $n$ or $m$ variables are not sufficient to allow access to all configurations of the system, if periodic boundary conditions are used.  However, MC moves that additionally flip adjacent pairs of spins are sufficient to allow exploration of the whole configuration space, as discussed in~\ref{app:mc}.

\section{Phase transition in the $2d$ spin model: behaviour in finite systems}
\label{sec:numerics}

The dFA model has a dynamical phase transition, which corresponds to a thermodynamic phase transition in the $2d$ spin model.  These mean that the functions $\hat{\cal G}(s),\hat{\cal I}(k)$ in (\ref{equ:hat-GG-II}) have the same behaviour as was illustrated in Fig.~\ref{fig:IG-bulk-sketch} for  ${\cal G},{\cal I}$.  This can be proven by direct generalisation of the arguments of~\cite{Garrahan2009}.

This Section presents numerical results for the thermodynamic transition in the $2d$ spin model.
As usual for numerical studies of phase transitions, the results are obtained in finite systems, so finite-size scaling analysis is required in order to interpret the results.  Since these systems are anisotropic, this analysis requires some care with the ratio of $Y/L$, in the thermodynamic limit~\cite{Privman1983}.
%We explain how the dependence of the behaviour on $Y/L$ can be understood by the theory of Privman and Fisher~
%We discuss general consequences of this theory for finite-size scaling at dynamical phase transitions and the interpretation of numerical results~\cite{Nemoto2017}.

\subsection{Dynamical phase transition}

To analyse the phase transition using data for finite systems we consider
\beq
k_{L,Y}(s) = \frac{1}{LY} \langle K \rangle_s = - \frac{\mathrm{d}}{\mathrm{d}s} \hat{G}_{L,Y}(s),
\eeq
which is the mean value of the order parameter, as well as the associated susceptibility
\beq
\chi_{L,Y}(s) =  - \frac{\mathrm{d}}{\mathrm{d}s} k_{L,Y}(s),
\eeq
which is proportional to the variance of $K$.  The system has an active phase with large $k$ and an inactive phase where most spins have $n_i=0$ and $k$ is small.  However, since configurations with $\sum_i n_{i,y}=0$ are forbidden, the inactive phase always includes at least one excitation line.

\begin{figure}
\includegraphics[width=155mm]{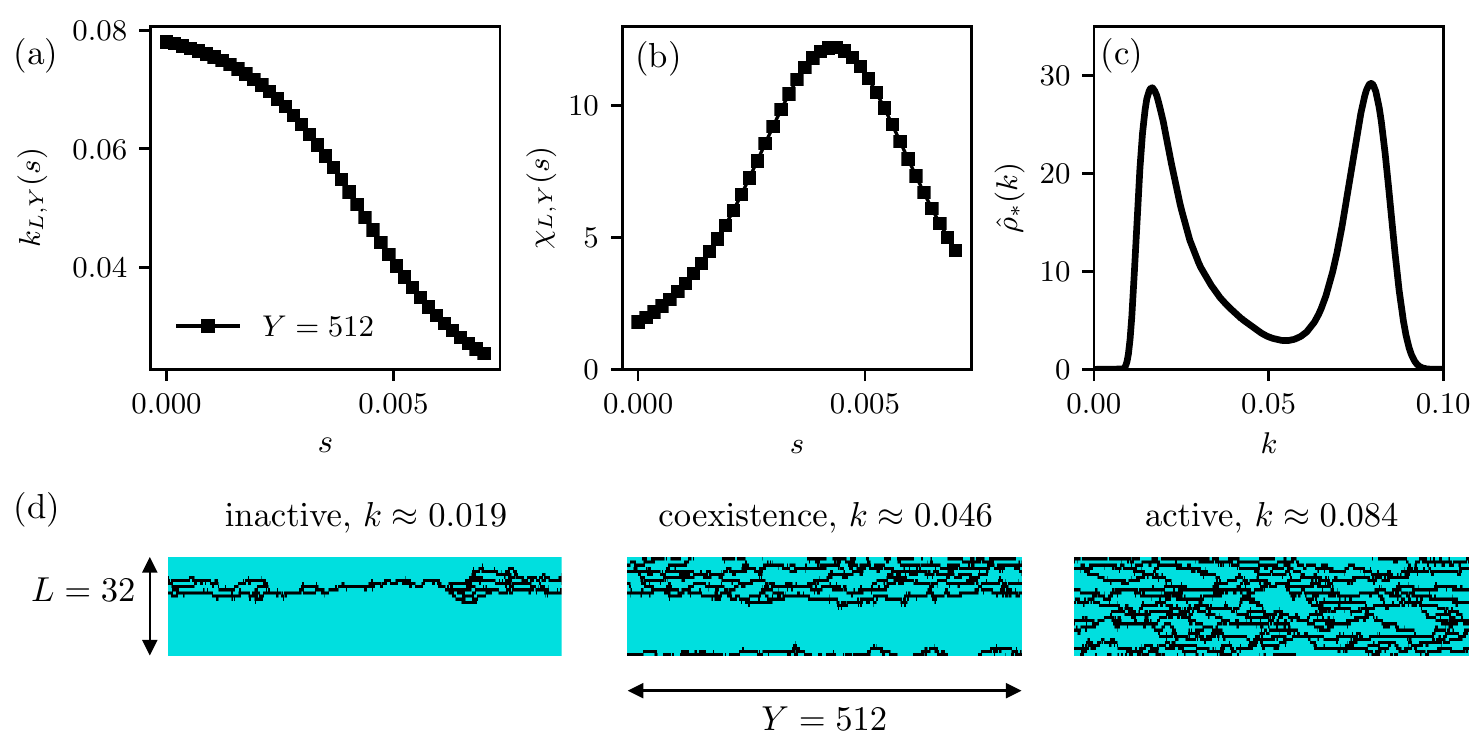}
\caption{Results for the $2d$ spin model $L=32$ and $Y=512$ and fully periodic boundary conditions.  (a) Order parameter $k(s)$.  (b) Susceptibility $\chi(s)$.  (c)~Histogram of the order parameter at $s=s^*_{L,Y}=0.0043$. (d) Representative configurations taken from the two peaks of the histogram (active/inactive phases) and from the trough (phase coexistence).  This is the expected behaviour in the regime $L\sim Y$.}
\label{fig:L32}
\end{figure}

Fig.~\ref{fig:L32} shows results for a system with $L=32$ and $Y=512$.  
As expected, the order parameter $k_{L,Y}$ shows a crossover as $s$ increases, with an associated peak in $\chi_{L,Y}$.  Define the field $s^*_{L,Y}$ to be the value of $s$ that maximises $\chi_{L,Y}(s)$.  Then the probability distribution 
\beq
\hat\rho_*(k) = \hat\rho_{L,Y}(k|s^*_{L,Y}) 
\eeq 
has a characteristic bimodal form in which the two peaks correspond to configurations from the active and inactive phases. The intermediate trough corresponds to configurations at phase coexistence.  The configurations that are typical of the trough of $\hat\rho_*$ depend on details of the system, including the aspect ratio $Y/L$ and the surface tension between the phases.  In this system, the fact that the inactive phase always contains at least one excitation line is also relevant.  For the case considered in Fig.~\ref{fig:L32}, one sees that phase coexistence corresponds to domains of active and inactive phases, with interfaces that lie parallel to the time-like axis.

\subsection{Finite-size scaling for $L\sim Y$}

The data of Fig.~\ref{fig:L32} show the classical features of a first order phase transition.  However, it is important to note that the behaviour of $\hat\rho$ and the nature of the representative configurations depend on the aspect ratio $Y/L$.   The behaviour of Fig.~\ref{fig:L32}(c,d) is expected when $L,Y$ are of the same order of magnitude, that is $L\sim Y$~\cite{Privman1983}.
In this case, the two peaks of $\hat{\rho}_*(k)$ appear at values $k_{\rm a},k_{\rm i}$ that are characteristic of the phases and depend weakly on $L,Y$.  Hence the variance of $k$ is of order unity, which corresponds to a peak in the susceptibility whose height diverges with the total size of the system:
\beq
\chi^* = \chi_{L,Y}(s^*_{L,Y})  \approx LY (k_{\rm a}-k_{\rm i})^2 / 4 \; .
\label{equ:chi-LY}
\eeq

Interfaces between active and inactive phases incur a free energy cost which depends on their orientation with respect to the lattice axes.  Hence we define two surface tension parameters $\Gamma_{\rm t},\Gamma_{\rm s}$, which are associated with interfaces that are parallel to the time-like and space-like axes respectively.  (See also~\cite{Jack2006jcp}.)
Since the interfaces in Fig.~\ref{fig:L32}(d) are parallel to the time-like axis, we infer that 
%these have the lowest cost: $Y \Gamma_{\rm t} < L \Gamma_{\rm s}$.  
%$\Gamma_{\rm t}$ is much smaller than $\Gamma_{\rm s}$: $\Gamma_{\rm t} < (L/Y) \Gamma_{\rm s}$ (with $L=32$, $Y=512$) \textbf{[Where ]}.
%
the costs of the time and space directions satisfy $Y \Gamma_{\rm t} \ll L \Gamma_{\rm s}$ so that  $\Gamma_{\rm t} \ll (L/Y) \Gamma_{\rm s}$.
The total length of the interface is $2Y$, so the peak-to-trough ratio of the distribution $\hat\rho_{L,Y}^*$ scales as $\ee^{-2Y\Gamma_{\rm t}}$, if one considers systems where the aspect ratio $Y/L$ is held constant as $Y\to\infty$.   With this scaling we emphasise that for values of $k$ within  the trough (that is, $k_{\rm i} < k < k_{\rm a}$), one has
\beq
\lim_{L,Y\to\infty} \frac{-1}{LY} \log \hat\rho_{L,Y}(k|s^*_{L,Y}) = \lim_{L,Y\to\infty} (2\Gamma_{\rm  s}/L) = 0
\label{equ:rho-subex}
\eeq
Hence from  (\ref{equ:hat-GG-II}) one has $\hat{\cal I}(k)=0$ throughout this range, consistent with Fig.~\ref{fig:IG-bulk-sketch}.
The physical content of this result is that, in terms of cost, differences between configurations from the trough and the peaks of $\hat{\rho}$ are localised at the interfaces. 
%
%\rlj{(T.N.: This sentence might be a bit confusing?: the difference between the white (or grey) configuration and the striped configuration in Fig.4(a) doesn't look like local.)}
%
Since the interfaces occupy a vanishing fraction of the system, the difference in free energy between these configurations is subextensive, and the value of $\frac{1}{LY}\log\hat\rho$ is the same.  This is related to the Maxwell construction and the double tangent construction in thermodynamics~\cite{Huang-intro}.

%Physically, the content of this formula is that the 
%interfacial free energy is subextensive and does not contribute to the bulk free energy density.
%This is related to the Maxwell construction in thermodynamics.  It means that values of $k$ between $k_{\rm i}$ and $k_{\rm a}$ can be achieved by configurations that contain large domains of the active and inactive phases, separated by interfacial regions that occupy a vanishingly small fraction of the system.  This means that the free energy of these configurations differs from that of the pure phases by a contribution that vanishes when divided by the (total) system size $LY$.
%(\emph{make the point here that 
%${\cal I}(k) = - s^* k - \lim_{L,Y\to\infty} \frac{1}{LY} \log \hat\rho^*_{L,Y}(k)$ so this means that ${\cal I}$ has a linear
%segment in this range of $k$.  Hence ${\cal I}''(k)=0$: it is non-concave but it is not strictly convex.})

\begin{figure}
\includegraphics[width=140mm]{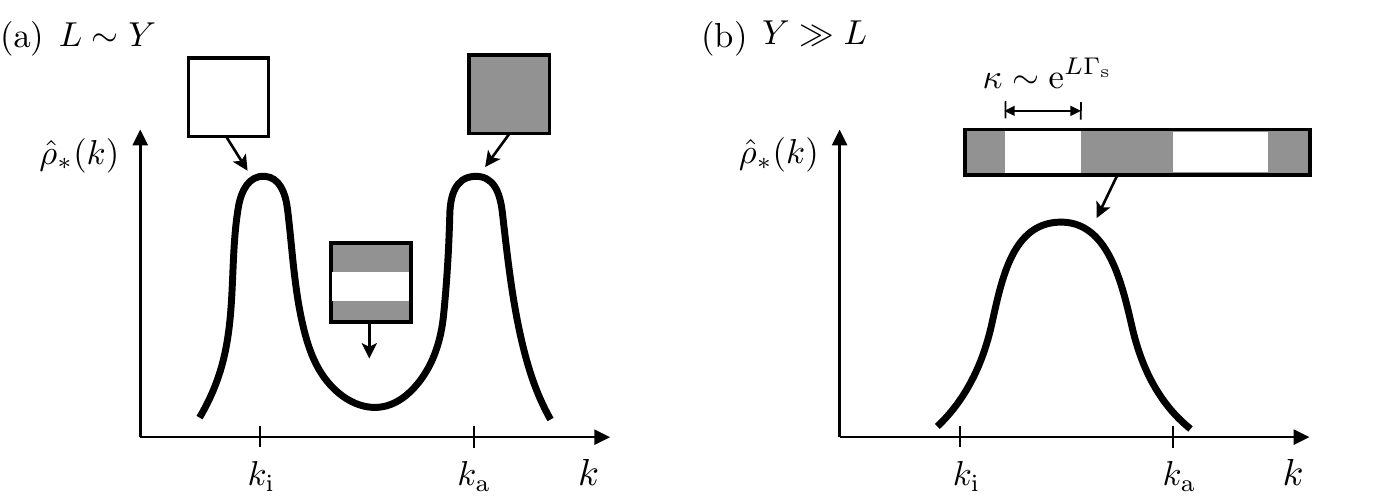}
\caption{The finite-size scaling crossover of Privman and Fisher~\cite{Privman1983}.  (a) For systems with $Y\sim L$, the classical picture of phase coexistence has a bimodal distribution of the order parameter.  Representative configurations of the system are illustrated.  The peaks correspond to the pure phases (illustrated in white and grey) and the trough corresponds to phase coexistence, shown here with domain walls parallel to the horizontal axis, as in Fig.~\ref{fig:L32}.
(b) For systems with $Y\gg L$ the situation is different.  The distribution of the order parameter is unimodal and typical configurations have many domains of both phases.  The typical distance between domain walls scales as $\ee^{L\Gamma_{\rm s}}$, see the text for a discussion.}
\label{fig:privman-sketch}
\end{figure}

\subsection{Finite-size scaling and consequences for dynamical large deviations}
\label{sec:privman}

So far, we analysed the behaviour of a finite system following the standard methods of thermodynamics.  To make contact with the methods used for dynamical phase transitions, recall from (\ref{equ:GL}) that the dynamical free energy is conventionally defined with a limit of large time $\tau$ at fixed system size $L$.  In the $2d$ spin model, this corresponds to taking $Y\to\infty$ at fixed $L$.  The finite-size scaling behaviour for $Y\gg L$ is quite different to that shown in Fig.~\ref{fig:L32}, which applies for $Y\sim L$.
This distinction was discussed by Privman and Fisher~\cite{Privman1983}, who analysed thermodynamic phase transitions for different aspect ratios $Y/L$.  This Section discusses the physical picture that follows from their analysis. Its consequences for the dynamical phase transition in the FA model will be discussed in Section~\ref{sec:linktoDPT}.

From general thermodynamic arguments, the free energy $\hat{\cal G}$ and its Legendre transform $\hat{\cal I}$ in (\ref{equ:hat-GG-II}) are bulk quantities, independent of the its aspect ratio.  In particular, one should obtain  the same result on interchanging the order of limits in (\ref{equ:hat-GG-II}), or on taking $L,Y\to\infty$ together, with a fixed ratio $Y/L$.  Moreover, Fig.~\ref{fig:IG-bulk-sketch} indicates that $\hat{\cal I}=0$ for $k<\overline{k}$ in the FA model.  Consistency of this picture with the dFA model requires that $\frac{1}{LY}\log\hat\rho_{L,Y}(k|s^*_{L,Y})\to0$ in large systems, throughout the range $0<k<\overline{k}$ [see Eq.~(\ref{equ:rho-subex})].
%
% \rlj{(T.N.: more explanations might be useful? The picture in Fig.~\ref{fig:IG-bulk-sketch} discusses $\frac{1}{LY}\log\hat\rho_{L,Y}(k|0)$ instead of $\frac{1}{LY}\log\hat\rho_{L,Y}(k|s^*)$. How is the property of $\frac{1}{LY}\log\hat\rho_{L,Y}(k|s^*)$ derived from the one of $\frac{1}{LY}\log\hat\rho_{L,Y}(k|0)$?)}
%
For the situation  
%this indicates that for large systems we must have $k_{\rm a}\to \overline{k}$ and $k_{\rm i}\to0$, and that $s^*_L\to0$, and that the peak-to-trough ratio of $\rho_*(k)$ does not diverge exponentially in $LY$.  
%In fact, 
where $L\sim Y$ and the aspect ratio $\alpha=Y/L$ is fixed in the thermodynamic limit, the expected behaviour is that
\beq
\log \hat \rho_{L,Y}(k|s^*_{L,Y}) \approx - L {\cal F}_{\sim}(k,\alpha,\Gamma_{\rm s},\Gamma_{\rm t}), \qquad k_{\rm i}<k< k_{\rm a}
\label{equ:surface-LDP}
\eeq
%\rlj{(T.N.: more explanations of why we suddenly change our target from $\hat \rho_{L,Y}(k|s^{*})$ to $ \hat \rho_{L,Y}(k|0)$ might be helpful?)}
where ${\cal F}_{\sim}$ is the free energy cost for the interfaces associated with phase coexistence.  The function ${\cal F}_{\sim}$ also depends on the boundary conditions. The general form (\ref{equ:surface-LDP}) is consistent with the behaviour of (\ref{equ:rho-subex}), which is relevant for the specific example in Fig.~\ref{fig:L32}(d): in that case $ {\cal F}_{\sim}(k,\alpha,\Gamma_{\rm s},\Gamma_{\rm t})=2\alpha\Gamma_{\rm t}$.

Note that (\ref{equ:surface-LDP}) is an LDP with speed $L$, which describes the behaviour of the system at phase coexistence.  Compared with the bulk result (\ref{equ:hat-GG-II}), it gives a more detailed description of the behaviour of $\hat\rho_{L,Y}$,  for ${k_{\rm i}}<k<k_{\rm a}$.  That is, the probability density $\hat\rho$ obeys two  LDPs: the ``bulk'' result (\ref{equ:hat-GG-II}) which has speed $LY$ and  the ``interfacial'' result (\ref{equ:surface-LDP}).   The bulk LDP has rate function zero throughout $0<k<\overline{k}$, and this is independent of aspect ratio and boundary conditions.  The interfacial LDP has a non-trivial rate function in this range of $k$; it depends on both the aspect ratio and the boundary conditions.

To make contact with dynamical phase transitions, we now consider systems with $Y\gg L$, which are relevant for the dynamical LDP (\ref{equ:ldp}) and for (\ref{equ:GG-II}).  Fig.~\ref{fig:privman-sketch} shows the behaviour of $\hat\rho_{L,Y}$ predicted by the theory of Privman and Fisher~\cite{Privman1983}, and corresponding configurations of the system.   For $Y\gg L$, phase coexistence is dominated by interfaces that are parallel to the space-like direction.  The free-energy cost of a single interface   is $\Gamma_{\rm s}L$ and these interfaces can appear anywhere along the time-like direction.  It follows that for very large $Y$, the statistical properties of these domain walls are those of a one-dimensional ideal gas with density 
$\ee^{-\Gamma_{\rm s}L}$.  Hence their number follows a Poisson distribution with mean
\beq
\overline{n} \approx Y \ee^{-\Gamma_{\rm s}L} \; .
\eeq
Recalling (\ref{equ:GL}) we are considering $Y\to\infty$ at fixed $L$: to saturate this limit requires $\overline{n}\gg 1$ or $Y\gg \ee^{\Gamma_{\rm s}L}$.   In this work, we use $Y\gg L$ as a short-hand for the limit where $Y\to\infty$ before any limit of large $L$.
This corresponds to an extremely large aspect ratio $Y/L$.  In thermodynamics, this regime is less often studied than the classical case $Y \sim L$, see however~\cite{Privman1983}.  It is also harder to access by conventional MC simulations.  However, it is the natural limit for the population dynamics (cloning) methods that are used to analyse dynamical large deviations~\cite{Giardina2006,Lecomte2007b}.

In this limit, the Poisson distribution for the number of domains means that the distribution $\hat\rho^*(k)$ is peaked at $(k_{\rm a}+k_{\rm i})/2$.  The probability to have $k\approx k_{\rm a}$ or $k\approx k_{\rm i}$ can be estimated as the probability to have no domain walls at all, which for a Poisson distribution is simply $\ee^{-\overline{n}}$.  In fact the statistics of the configurations shown in Fig.~\ref{fig:privman-sketch}(b) are exactly those of a \emph{one-dimensional} Ising model of size $Y$ where the density of domain walls is $ \ee^{-\Gamma_{\rm s}L} $.  Equivalently one can identify these configurations with trajectories of a two-state Markov chain where each state corresponds to one of the phases, and the rate for transitions between the phases is $\ee^{-\Gamma_{\rm s}L}$.  In either case, the statistics of the activity are easily computed (see \ref{app:1d}) and one obtains
\beq
\log \hat\rho_{L,Y}(k|s^*_{L,Y}) \approx - Y {\cal F}_{\gg}(k), \qquad k_{\rm i}<k< k_{\rm a}
\label{equ:LDP-gg}
\eeq
with
\beq
{\cal F}_{\gg}(k) =  \ee^{-\Gamma_{\rm s}L} \,f\!\left(\frac{2(k-k_0)}{\Delta k}\right) , \qquad f(x) = 1-\sqrt{1-x^2} \; ,
\label{equ:triv-ldp}
\eeq
where $k_0=(k_{\rm a}+k_{\rm i})/2$ is the average activity of the two phases and $\Delta k = k_{\rm a}-k_{\rm i}$ is their difference.
Eq.~(\ref{equ:LDP-gg}) is the analogue of (\ref{equ:surface-LDP}) for systems with $Y\gg L$.    We identify it as an LDP with speed $Y$.  Again, this result gives extra detail on the behaviour of $\hat\rho_{L,Y}$, in the regime of phase coexistence.  Consistent with the bulk LDP one has $\frac{1}{LY}\log\rho\to0$.  
Note however that the LDP (\ref{equ:LDP-gg}) is quite different from its counterpart (\ref{equ:surface-LDP}) for $L\sim Y$.  In particular, the rate function ${\cal F}_\gg$ has a simple form that is convex, consistent with the unimodal distribution $\hat\rho_*$ in Fig.~\ref{fig:privman-sketch}(b).  In contrast, ${\cal F}_\sim$ in  (\ref{equ:surface-LDP}) is non-convex in general, consistent with the bimodal distributions in Figs.~\ref{fig:L32}(c) and \ref{fig:privman-sketch}(a).    

%Bimodal distributions similar to Fig.~\ref{fig:privman-sketch} have been used as evidence for first-order dynamical phase transition, for example in~\cite{Hedges2009,Elmatad2010} -- those works used transition path sampling and work in the regime corresponding to $L\sim Y$.  However, for systems with $Y\gg L$ then we emphasise that unimodal behaviour for $\hat\rho_*$ is fully consistent with the existence of a phase transition, despite some of the arguments presented in~\cite{Whitelam2018slow}.

\begin{figure}
\includegraphics[width=150mm]{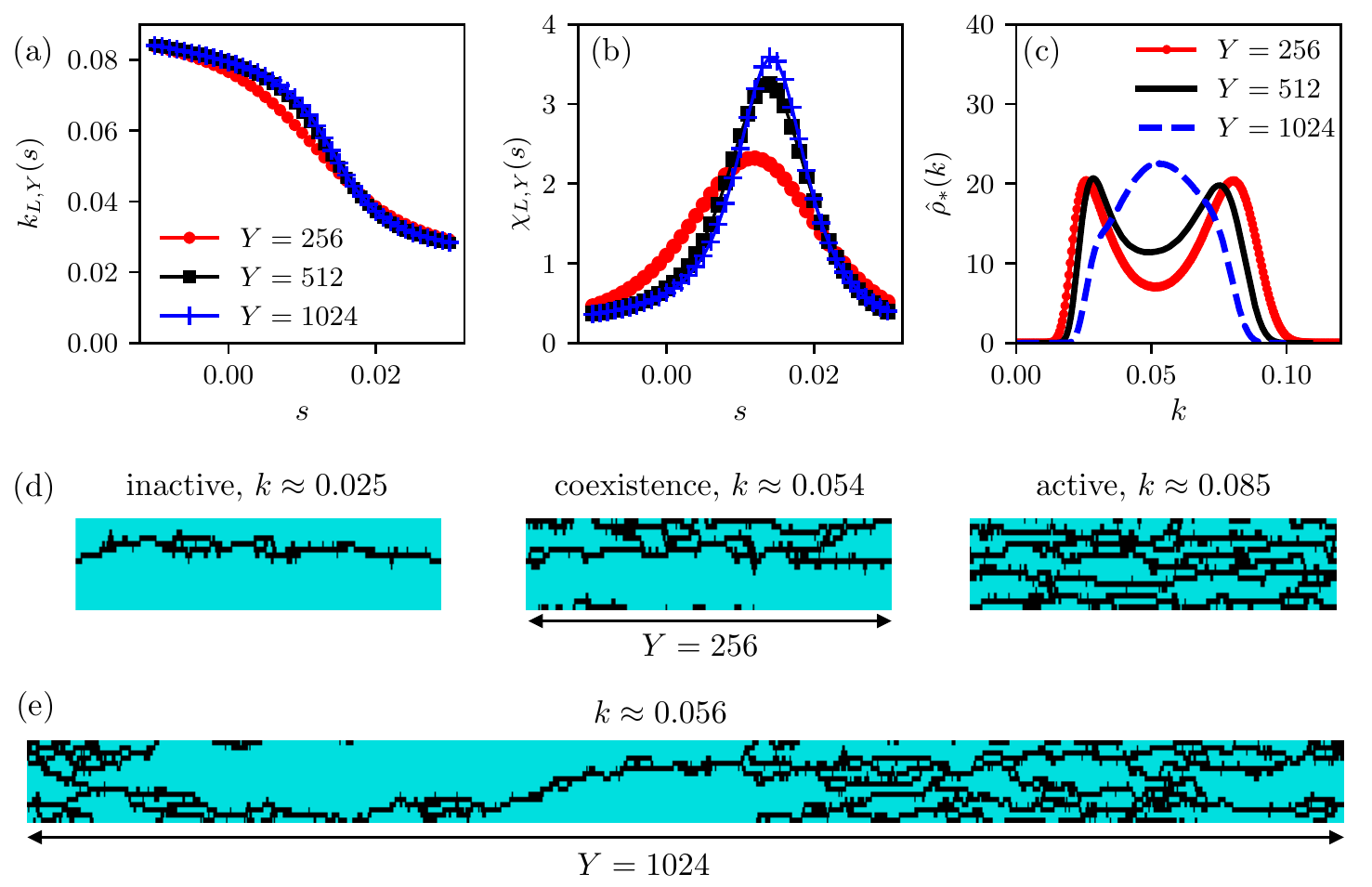}
\caption{Results for $L=16$.  (a) Activity $k(s)$ for different values of $Y$, as labelled, the data for $Y=512$ are very close to those for $Y=1024$. (b) Susceptibility $\chi(s)$, the symbols and colours are the same as for panel (a).  (c)~Histograms of the activity at $s=s^*$.  For smaller $Y$ one sees a bimodal structure but for larger $Y$ one sees a single peak, this is the Privman-Fisher crossover~\cite{Privman1983}.  (d)~Configurations at $Y=256$ showing behaviour that is representative of the two peaks in $\rho{*}$ (inactive/active phases), and of the intervening trough.
(e) A configuration from $Y=1024$ that is representative of the peak of $\rho_*$, showing both phases coexisting in a single configuration.   For even larger values of $Y$, typical trajectories would include multiple domains of each phase.
}
\label{fig:L16}
\end{figure}

Noting that the LDP (\ref{equ:LDP-gg}) applies to the distribution $\hat\rho_{L,Y}(k|s^*_{L,Y})$, define $s(k)=s^*_{L,Y}-{\cal F}_{\gg}'(k)$, and note that the inverse of the function $s(k)$ is $k_{L,\infty}(s)$.  Hence 
\beq
k_{L,\infty}(s) = k_0 -  \frac{(s-s^*_{L,\infty})(\Delta k)^2}{4\sqrt{(s-s^*_{L,\infty})^2(\Delta k)^2 + (\ee^{-2\Gamma_{\rm s}L}/4) }}
\label{equ:k-first}
\eeq
which describes the crossover between the phases as shown (for example) in Fig.~\ref{fig:L16}(a).
We will return to this result in Sec.~\ref{sec:interfacial}, below.
Note that (\ref{equ:LDP-gg}) and (\ref{equ:k-first}) are generic for first-order dynamical phase transitions in two dimensions.  The extension to higher dimensions is straightforward in the case where the system size along one dimension is much larger than the size along all others, in which case one should replace $\Gamma_{\rm s}L$ by $\Gamma_{\rm s}L^{d-1}$ for a suitably-defined surface tension  parameter $\Gamma_{\rm s}$.  For dynamical phase transitions of mean-field systems, one expects a similar formula with an ``interfacial cost'' $\Gamma_{\rm s}N$ where $N$ is the system size, see~\cite{Nemoto2014jstat}.
In diffusive systems described by the Macroscopic Fluctuation Theory, see~\cite{Baek_2018} for a discussion of similar phenomena.
%
%Here, ${\cal F}_{\gg}(k)$ is a function that vanishes at $k=(k_{\rm a}+k_{\rm i})/2$ and is equal to unity at $k=k_{\rm a},k_{\rm i}$.  
%This function can be obtained using the ideal gas statistics of the domain walls.  \emph{do this!}

%The total interfacial free energy cost of a system with $n$ domains is approximately $nL\Gamma_L$ (for every domain there is an interface parallel to the space-like axis, of length $L$).  Since $L$ is finite then so is this free energy cost, and one expects a finite density of these domain walls (along the time-like axis). Their number will be approximately $\overline{n} \sim Y\ee^{-L\Gamma_L}$ and the typical size of such a domain (measured along the time-like axis) is approximately

\subsection{$2d$ spin model for $Y\gg L$}
\label{sec:YggL}

Fig.~\ref{fig:L16} illustrates the finite-size scaling crossover for the $2d$ spin model with $L=16$.  If $Y$ is not too large then one observes a bimodal distribution of the order parameter, as in Fig.~\ref{fig:L32}.  However, one sees for larger $Y$ a crossover to a unimodal form, consistent with Fig.~\ref{fig:privman-sketch}.  The representative configurations also show the crossover anticipated in Fig.~\ref{fig:privman-sketch}, although  the typical configurations have only one active and one inactive domain, even for the largest $Y$ that we analysed.  (To observe many domains would require even larger $Y$, which is numerically expensive.)

For large $Y$, one also sees that the order parameter $k_{L,Y}(s)$ approaches a limiting form $k_{L,\infty}(s)$.  To understand this, note that the thermodynamic analogue of the dynamical SCGF $G_L(s)$ in (\ref{equ:GL}) is 
\beq
\hat{G}_{L,\infty}(s) = \lim_{Y\to\infty} \hat{G}_{L,Y}(s) \; .
\eeq 
(To be precise, this corresponds to $G_L(s)/L$.)  Following the same arguments as in the dynamical case, this limit exists and $\hat{G}_{L,\infty}$ is strictly convex and analytic.  Its derivative gives the limiting form of the order parameter $k_{L,\infty}(s) =  -G'_{L,\infty}(s)$.  From  (\ref{equ:k-first}) one has also the corresponding susceptibility which  is
\beq
\chi_{L,\infty}(s^*) \approx  \frac{(\Delta  k)^2}{4}  \ee^{L \Gamma_{\rm s}}  \; .
\label{equ:chi-kappa}
\eeq
This quantity diverges exponentially with $L$, in contrast to the power-law divergence of (\ref{equ:chi-LY}), which we recall was applicable for $L\sim Y$.  This exponential scaling was observed for the (original) FA model in~\cite{Nemoto2017}, where the connection to the work of Privman and Fisher was identified.  Quantitative predictions for $\Gamma_{\rm s}$ were also obtained in that case, see Sec.~\ref{sec:interfacial}.

%Finally, we also also consider the distribution $\hat\rho_{L,Y}^*(k)$ in this limit.  The variance of this distribution scales as $Y/\kappa_L$ which we recall is small compared to unity (because we are in the regime $Y\gg L$).  However, this variance is large enough that the curvature of ${\cal I}(k)$ can be shown to vanish throughout the range $k_{\rm i}<k<k_{\rm a}$.  Hence one recovers the result (Maxwell construction) that was previously obtained from the regime $L\sim Y$, that ${\cal I}(k)$ is linear throughout this range.

\subsection{Connection to previous studies of dynamical phase transitions}
\label{sec:linktoDPT}

We have used the $2d$ spin model of Sec.~\ref{sec:spin2d-def}  to illustrate the finite-size scaling crossover of Privman and Fisher~\cite{Privman1983}.  Since the thermodynamics of this model are directly related to dynamical large deviations of the dFA model (and the FA model), we are now able to rationalise some of the previous work on finite-size scaling of dynamical phase transitions. 

In particular, note that analysis of these phase transitions by transition path sampling~\cite{Bolhuis2002} typically focusses on finite-size scaling with $L\sim Y$~\cite{Hedges2009,Elmatad2010}. 
For example the analysis of~\cite{Hedges2009} assumes the scenario shown in Fig.~\ref{fig:L32}, with a bimodal histogram of the order parameter.  The associated peak-to-trough ratio increases with $Y$ (or in that case $\tau$), which was used in~\cite{Hedges2009} as evidence for a first-order transition.  Similar behaviour is seen in~\cite{Elmatad2010,SpeckChandler2012}.  In those cases one expects a power-law divergence of $\chi_{L,Y}(s^*)$.

On the other hand, analyses of dynamical  phase transitions~\cite{Giardina2006,Lecomte2007b} by cloning methods typically concentrate on the limit $Y\gg L$, as do analytical (and other) methods that estimate of the largest eigenvalue of a tilted generator~\cite{banuls2019}.  In these cases the behaviour of $\hat\rho_*$ close to its peak is described by~(\ref{equ:k-first}), in which ${\cal F}_\gg$ is a convex function.  It follows that $\hat\rho_*$ is unimodal, consistent with Figs.~\ref{fig:privman-sketch}(b) and \ref{fig:L16}(c).  We emphasise that this unimodal behaviour of $\hat\rho_*$ is fully consistent with the existence of a phase transition, despite some of the arguments presented in~\cite{Whitelam2018slow}.   In this case the susceptibility $\chi_{L,\infty}(s^*)$ is expected to diverge exponentially with $L$, as observed in Ref.~\cite{Nemoto2017} for the FA model.

%approach is typically to estimate the SCGF $G_L$ of (\ref{equ:GL}), which corresponds in the thermodynamic case to $\hat{G}_{L,\infty}$.  This is necessarily a convex function.  In this case one does not expect a bimodal histogram of the order parameter and the susceptibility $\chi_{L,\infty}(s^*)$ is expected to diverge exponentially with $L$.  Such an exponential divergence may be difficult to characterise numerically, but it was observed in~\cite{Nemoto2017} for the FA model.

%\fi

\section{Interfacial model of phase coexistence}
\label{sec:interfacial}

We have combined general theoretical arguments about first-order phase transitions with numerical calculations for the $2d$ spin model.  We now take an analytic approach to the dynamical phase transition in the FA model.  Specifically,  we expand on the work of~\cite{Nemoto2017}, and we also correct some typographical errors in that work.\footnote{%
In particular, the main formula (9) of reference \cite{Nemoto2017} is incorrect: $F(c)$ in the left hand side should be replaced by $F(c)/2$ and also the definition of $F(c)$ should be $(1/2)\log \left [ 1/(1-c) \right ]$ instead of $(1/2)\log \left [ c/(1-c) \right ].$}  
Our results yield a detailed understanding of the phase coexistence regime presented in the last section, and it gives a numerical value of the surface tension parameter $\Gamma_{\rm s}$ that determines the rate of exponential growth of the susceptibility.

\subsection{Definition of interfacial model}

We describe the phase transition in the FA model by a simplified interfacial model~\cite{Bodineau2012cmp,Bodineau2012jsp}, see also~\cite{Nemoto2017,Dolezal2019}.  We
assume that the FA model at time $t$ has a single large inactive domain where all spins have $n_i=0$.  Following~\cite{Bodineau2012cmp,Bodineau2012jsp}, we consider a continuous time Markov chain where the size of the \emph{active} domain at time $t$ is $x_t\in \{1,2,\dots,L\}$, so the size of the inactive domain is $L-x_t$.  The dynamical rule is that $x$ follows an asymmetric random walk: it increases by a step of $1$ with rate $2q$ and decreases by a step of $1$ with rate $2p$.  If such a jump would lead to $x>L$ or $x<1$ then it is rejected and $x$ is unchanged.  To reproduce the behaviour of the FA model requires
\beq
p = c(1-c), \qquad q = c \; .
\eeq
Since $p<q$ then the random walk is asymmetric and biased towards $x=1$ (which is the active phase).  To parameterise the asymmetry of the walk we define 
\beq
F=\frac12 \log \frac{q}{p}
\eeq
which is positive but small in magnitude for our examples (of order $c$).  The natural microscopic time scale in the problem is $(p+q)$.

For a trajectory of the FA model where the size of the active domain is $x(t)$, one expects a dynamical activity 
\beq
K(\tau) \approx \mathcal{K}(\tau)  = \overline{k} \int_0^\tau x(t) \mathrm{d}t
\label{equ:K-fa-interf}
\eeq
where the second equality defines ${\cal K}(\tau)$, which  is
the dynamical activity of a trajectory of the interfacial model.
Eq.~(\ref{equ:K-fa-interf}) assumes that behaviour within the active domain is close to the typical behaviour of the model, so the activity density there is close to $\overline{k}$. Note that $\bar k = 4c^2(1-c)$ in this model \cite{Nemoto2017,Bodineau2012cmp,Bodineau2012jsp}, different from the one introduced in previous sections.

For large $L$, it is convenient to replace the integer-valued domain size $x$ by the fraction of the system that is covered by this domain, that is 
\beq
y(t) = \frac{x(t)}{L}
\eeq
In~\cite{Nemoto2017}, this variable $y(t)$ was denoted by $\tilde x(t)$.  
The analogue of the SCGF $G_{L,\tau}(s)$ for this model is $\Psi_{L,\tau}(sL)$ where 
\beq
\Psi_{L,\tau}(\lambda) = \frac{1}{\tau} \log \left\langle {\rm e}^{-\lambda \int_0^\tau y(t) \mathrm{d}t} \right\rangle .
\label{equ:Psi-def}
\eeq
To the extent that the interfacial model captures the relevant physics in the FA model, we expect
\beq
G_{L,\tau}(s) \approx \Psi_{L,\tau}(sL).
\label{equ:G-Psi}
\eeq

The interfacial model does not capture all aspects of the FA model, in particular the fact that there are no fluctuations within the active phase is a coarse approximation.  However, we will find that it is valid (at best) in regimes where the trajectories of the FA model contain large coexisting domains of active and inactive phases, with at most one large domain being present at any time $t$.  This corresponds to values of $s$ that  are close to peaks of the susceptibility $\chi$ that can be seen in Figs.~\ref{fig:L32}, \ref{fig:L16}.

\subsection{Large deviation analysis}

\begin{figure}
\hspace{25mm}
\includegraphics[width=8cm]{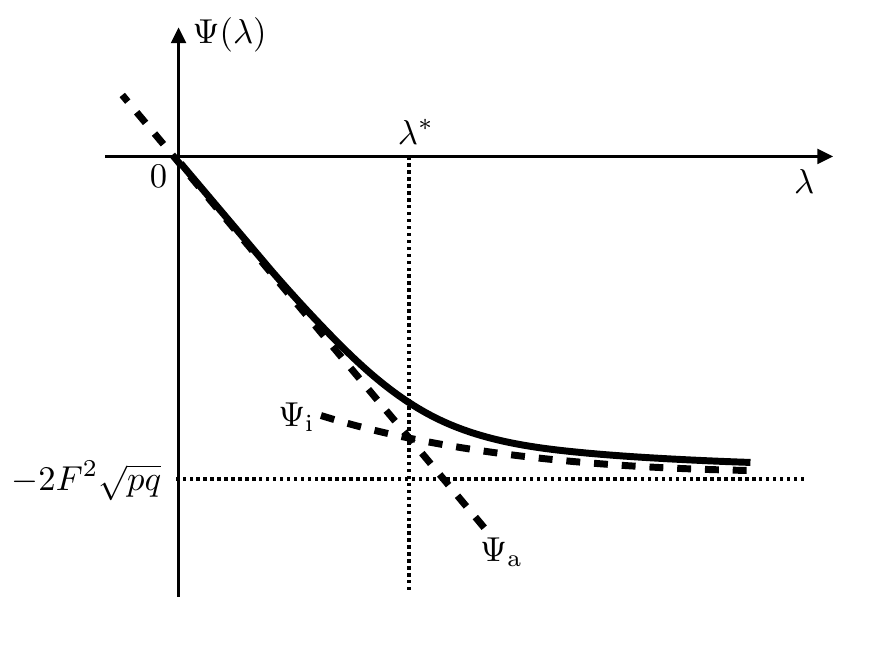}
\caption{Sketch of the SCGF $\Psi_L(\lambda)$, shown as a solid line.  The dashed lines show the behaviour of the active and inactive phases. The active phase has $\Psi_{\rm a}\approx -\lambda\overline{k}$ with a correction at order $1/L$ (the mean activity ${\cal K}/(L\tau)$ of this phase is $\overline{k}$).  Also $\Psi_{\rm i}\approx -2\sqrt{pq}F^2$, with a finite-size correction of order $L^{-2/3}$, see (\ref{equ:Psi-i}).  The dashed lines cross at $\lambda=\lambda^*\approx 2F^2\sqrt{pq}/\overline{k}$.  The SCGF $\Psi$ deviates from $\mathrm{max}(\Psi_{\rm a},\Psi_{\rm i})$ by a quantity that scales as $\ee^{-L\Gamma}$, where the constant $\Gamma$ is given by (\ref{equ:chi*-predict}).  This means that the curvature $\Psi''(\lambda)$ is exponentially large for $\lambda\approx\lambda^*$.}
\label{fig:fss-sketch}
\end{figure}

We now consider large deviations of the interfacial model in the limit of large $\tau$, so we define
\beq
\Psi_L(\lambda) = \lim_{\tau\to\infty} \Psi_{L,\tau}(\lambda)
\label{equ:PsiL}
\eeq
analogous to $G_L(s)$ in the FA model,  see (\ref{equ:GL}).  The following analysis applies for $\lambda\geq0$
and we also assume that $F\ll 1$, which simplifies the analysis.
The results are summarised in Fig.~\ref{fig:fss-sketch}.  For large-$L$ this result is consistent with Fig.~\ref{fig:IG-bulk-sketch}(b), but we emphasise again that the behaviour  for finite $L$ is specific to a system where the limit of large time $\tau$ is taken before any limit of large $L$.
Note also that $\Psi_L$ is decreasing and convex, so we have for $\lambda\geq0$ that
\beq
-\lambda\overline{k} \leq \Psi_L(\lambda) \leq 0 \; .
\label{equ:Psi-convex}
\eeq

\subsubsection{Eigenvalue problem in continuous space}
\label{subsec:eval-cts}

The SCGF can be obtained by solving an eigenproblem, based on the master equation for the dynamics \cite{Garrahan2009}.  Within the bulk of the system ($y\neq\tfrac1L,1$) one has
\beq
\fl\quad \Psi_L(\lambda) P_L(y) = 2p [ P_L(y+\tfrac1L) - P_L(y) ] + 2q [ P_L(y-\tfrac1L) - P_L(y) ] - \lambda y  \overline{k} P_L(y)
\label{equ:mat-bulk}
\eeq
where the eigenvector $P_L$ is a (discrete) probability distribution for $y$.  In  the following we sometimes omit the subscripts $L$, for compactness of notation.
At the boundaries one has 
\begin{eqnarray}
\Psi P(\tfrac1L) &=&  2p P(\tfrac2L) - 2q P(\tfrac1L) - \lambda\overline{k}/L
\nonumber\\ 
\Psi P(1) &=&  2q P(\tfrac{L-1}{L}) - 2p P(1) - \lambda\overline{k} \; .
\label{equ:mat-bdy}
\end{eqnarray}
Eqs.~(\ref{equ:mat-bulk}, \ref{equ:mat-bdy}) define
an eigenvalue problem for an $L\times L$ matrix, which is easily solved numerically.    Here we adopt an analytical approach, which gives insight into the limit of large $L$.  Our analysis also requires that $F\ll1$.

In the large-$L$ limit, we find that the eigenvector $P$ takes the form
\beq
P(y) = \exp[ Lf(y) ],
\label{equ:PLf}
\eeq
where $f$ is of order unity.  
%Physically, this equation means that the \emph{instantaneous} position $y_t$ of the interface obeys an LDP with speed $L$.  
For example, in the absence of any bias, $f(y)=2Fy$ and the the distribution of $y$ is exponentially increasing, with a maximum at $y=1$.  (This is the active phase.)  

\begin{figure}
\begin{center}
\includegraphics[width=150mm]{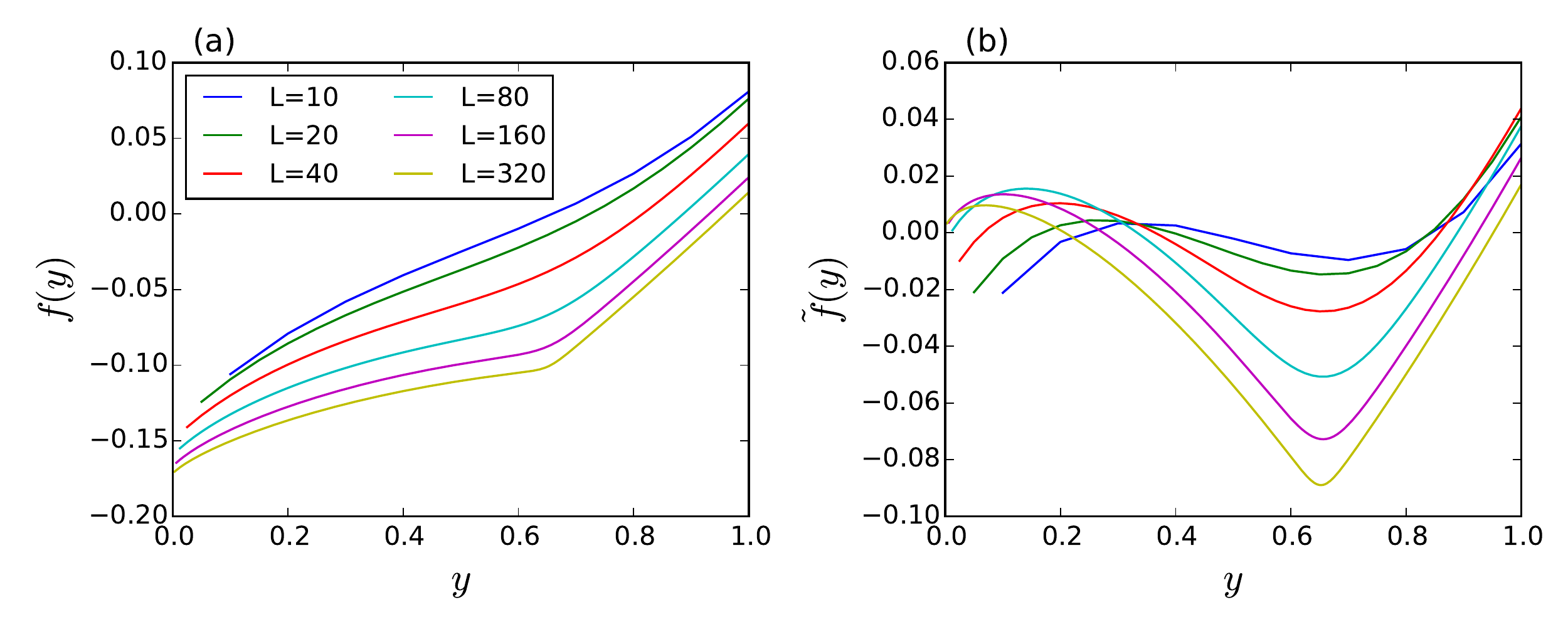}
\caption{The rate functions $f(y)$ and $\tilde f(y)$ at coexistence, evaluated for finite values of $L$ by diagonalizing the eigenvalue problem (\ref{equ:mat-bulk}). The value of $c$ is set to 0.3. The value of $\lambda$ is tuned to the special value that maximizes the curvature $\Psi^{\prime \prime}(\lambda)$ for each $L$. The figures show that $\tilde f(y)$ has two maxima at coexistence, whereas $f(y)$ does not.}
\label{fig:f_tildef}
\end{center}
\end{figure}

In general, the eigenfunction $P$ gives the distribution of the size of the active domain, evaluated at the final time $\tau$.
The meaning of (\ref{equ:PLf}) is that this size obeys an LDP with speed $L$ and rate function $-f(y)$.
It is convenient to define also $Q(y)=P(y)\ee^{-LFy}$, in which case $Q(y)^2$ is a (non-normalised) probability distribution for the domain size $y_t$, where $t$ is in the ``bulk'' of the trajectory (\emph{i.e.}~far from initial and final times).   In the notation of~\cite{Nemoto2017}, we thus have $P_{\rm ave}(y)\propto Q(y)^2$ and $P_{\rm end}(y)=P(y)$.  One sees that  $P_{\rm ave}$ also satisfies an LDP with speed $L$, that is
\beq
Q(y)^2 = \exp[ L \tilde{f}(y)] \; , \qquad \tilde{f}(y) = 2[ f(y)-yF ]  \; .
\label{equ:QLf}
\eeq
% and rate function $2(yF-f(y))$.
%
At the dynamical phase transition, $\tilde{f}(y)$ presents two  maxima as shown in Fig.~\ref{fig:f_tildef} for finite $L$, and the corresponding values of $y$ are the sizes of the active domain in the coexisting phases.
%.  For $\tilde{f}$, the maxima are degenerate at the phase coexistence point, and they appear at the mean values of $y$ in the coexisting phases.\footnote{%
%\color{red}Technically, this $f(y)$ minus $F y$ [{\it i.e.}, $\log Q(y)/L$ where $Q(y)$ is introduced above Eq.(\ref{equ:mu-psi})] corresponds to the large deviation function of $y$ in the system biased by the activity ${\cal K}$. So $f(y)$ itself does not have the two minima at the dynamical phase transition. See numerical examples of $f(y)$ in Fig.~\ref{fig:continuous_exact}(b).}
We emphasise however that the LDPs (\ref{equ:PLf}, \ref{equ:QLf}) for $y_t$ are distinct from the LDP for the time-integrated quantity ${\cal K}$ that is our main interest here.

The passage from the discrete-space problem to a continuum one is discussed in~\ref{app:continuum}.  
We define
\beq
\mu = \left( \frac{\lambda\overline{k}}{2\sqrt{pq}} \right)^{1/3}, \qquad \psi_L = \frac{ \Psi_L(\lambda) }{2 \sqrt{pq} } + F^2 \; .
\label{equ:mu-psi}
\eeq
Note that $\mu\geq0$ is known (it depends only on the parameters of the model) but $\psi_L$ is unknown (it depends on the eigenvalue $\Psi_L$).  We recognise $\psi_L$ as the difference between the solid line and the horizontal dotted line in Fig.~\ref{fig:fss-sketch}.
The function $Q$ satisfies the self-adjoint eigenvalue equation (\ref{equ:Q-eigen-F}) which reduces to
\beq
\frac{1}{L^2} Q''_L(y) - \mu^3 y Q_L(y) = \psi_L Q_L(y)
\label{equ:airy-Q}
\eeq
This equation is to be solved subject to the boundary conditions (\ref{equ:bdy-Q-F}) which are
\beq 
\fl\qquad
\frac{Q'(0)}{LQ(0)} =  \gamma_0  = F+ (\psi-F^2)  + O(F^3) , \quad \frac{Q'(1)}{LQ(1)} =  \gamma_1  =  F+(\psi+\mu^3-F^2)   + O(F^3) .\;
\label{equ:bdy-Q-psi}
\eeq
This equation defines $\gamma_0,\gamma_1$. (The quantities in round brackets are $O(F^2)$, we retain them for later convenience.
Note that ignoring terms at $O(F^3)$ can sometimes lead to artificial solutions to the eigenproblem with $\psi>0$, see \ref{Appendix:numerical_demonstrations}.)
% We thus add a condition that we look for the largest {\it negative} $\psi$. See \ref{Appendix:numerical_demonstrations} for more detail.) 

Since the coefficient of $Q''$ in (\ref{equ:airy-Q}) is small, one may obtain approximate solutions to the eigenproblem.  This is equivalent to a WKB-like saddle-point analysis method.  Here we take a different route, using that (\ref{equ:airy-Q}) can be solved exactly using Airy functions.

\subsubsection{Solution of the eigenvalue problem}

The Airy functions are solutions of the Airy equation $f''(x)-xf(x)=0$, and are denoted by $\Ai(x)$ and $\Bi(x)$.  
For negative $x$, these functions oscillate around zero.  For positive $x$, $\Ai$ is a decreasing function and $\Bi$ is increasing.  For large positive $x$ they behave as 
\begin{eqnarray}
\Ai(x) &\simeq& \exp\left(-\tfrac23 x^{3/2}\right) \frac{1}{2\sqrt{\pi}x^{1/4}} \left[ 1 + O(x^{-3/2}) \right] \; ,
\nonumber\\
\Bi(x) &\simeq& \exp\left(\tfrac23 x^{3/2}\right) \frac{1}{2\sqrt{\pi}x^{1/4}} \left[ 1 + O(x^{-3/2}) \right] \; .
\label{equ:AiBi-exp}
\end{eqnarray}
Note in particular that $\Bi(x)$ grows super-exponentially when its argument is large.

By changing variables in (\ref{equ:airy-Q}), its solution can be seen to be
\beq
Q_L(y) = \hata_L \Ac_L(y) + \hatb_L \Bc_L(y)
\label{equ:QL}
\eeq
where $\hata_L,\hatb_L$ are coefficients with a normalisation condition $\hata_L+\hatb_L=1$ and  
\begin{eqnarray}
\Ac_L(y) &=& \Ai\left[ L^{2/3} \left(\mu y + \frac{\psi_L}{\mu^2} \right) \right], 
\nonumber\\
\Bc_L(y) &=& \frac{1}{\Bi\left( L^{2/3} \left(\mu + \frac{\psi_L}{\mu^2} \right) \right)} \Bi\left[ L^{2/3} \left(\mu y + \frac{\psi_L}{\mu^2} \right) \right]. 
\end{eqnarray}
Note that $\Bc$ is normalised such that $\Bc_L(1)=1$.  From (\ref{equ:Psi-convex}, \ref{equ:mu-psi}) we have
\beq
-\mu^3 \leq (\psi - F^2) \leq 0
\eeq
which implies in particular that $\psi+\mu^3$ is (strictly) positive, and the asymptotic expansions of (\ref{equ:AiBi-exp}) are relevant.
To keep track of exponential factors in the large-$L$ limit, it is useful to define 
\beq
\eps_L = \exp\left[ -\frac{2}{3} L \left( \mu + \frac{\psi_L}{\mu^2} \right)^{3/2} \right] \; .
\label{equ:eps}
\eeq 
which is exponentially small when $L$ is large.
With this choice one sees that $\Ac_L(1)$ and $\Bc_L(0)$ are both of order $\eps_L\cdot L^{-1/6}$.

The physical interpretation of (\ref{equ:QL}) is that $\Ac_L$ and $\Bc_L$ have peaks near $0$ and $1$ respectively and represent contributions to the eigenfunction from the two phases.  The weights $\hata,\hatb$ and the unknown quantity $\psi_L$ must now be determined from the boundary condition (\ref{equ:bdy-Q-psi}).   
This requires that 
\beq
\frac{\hata_L}{\hatb_L} = \frac{ \Bc'(0) - L \gamma_0 \Bc(0) }{ L \gamma_0 \Ac(0) - \Ac'(0) } 
= \frac{ \Bc'(1) - L \gamma_1 \Bc(1) } { L \gamma_1 \Ac(1) - \Ac'(1) }
\label{equ:ab-1}
\eeq
Recall that $\Bc(0)$ and its derivative are exponentially small when $L$ is large, and similarly for $\Ac(1)$.  To compensate these exponential factors, define 
\beq
\eta_L = [ \Bc'(0) - L \gamma_0 \Bc(0)] /  \eps_L, \qquad \xi_L=[ L \gamma_1 \Ac(1) - \Ac'(1)]/\eps_L \;.
\eeq  
To simplify notation define also 
\beq
V_L=L \gamma_0 \Ac(0) - \Ac'(0), \qquad W_L=\Bc'(1) - L \gamma_0 \Bc(1) \; .
\eeq
Then the second equality in (\ref{equ:ab-1}) is
\beq
\frac{\eps_L\eta_L }{V_L} = \frac{W_L}{\eps_L\xi_L } \; .
\label{equ:VW}
\eeq
The objects $\eta_L,V_L,W_L,\xi_L$ 
do not depend on $\hata_L$ or $\hatb_L$ but they do depend on $\mu,F$ (which are parameters of the model) and on $\psi_L$.  The presence of the small quantity $\eps_L$ in these equations allows the behaviour of $\psi_L$ to be determined.

Observe that none of $\eta_L,V_L,W_L,\xi_L$ grow exponentially with $L$. 
Hence there are three possibilities in (\ref{equ:VW}).
The first is that $W_L$ is proportional to $\eps_L^2$ for large $L$, but $V_L$ is well-behaved (in the sense that it does not have an exponential dependence on $L$).  In this case $\hata_L/\hatb_L$ is proportional to $\eps_L$ so the eigenfunction $Q$ is dominated by a peak at $y\approx 1$.  This will correspond to the active phase.
The second possibility is that $V_L$ is proportional to $\eps_L^2$ for large $L$, but $W_L$ is well-behaved.  In this case $\hatb_L/\hata_L$ is proportional to $\eps_L$ so the eigenfunction $Q$ is dominated by a peak at $y\approx 0$.  This will be the inactive phase.  
The third possibility is that both $V_L$ and $W_L$ are proportional to $\eps_L$ so that $\hatb_L/\hata_L$ is well-behaved.  This corresponds to phase coexistence. 

We consider the three cases in turn.  Note there is no symmetry between the two phases in this problem.

\subsubsection{Active phase}

The active phase corresponds to exponentially small values of $W_L$ which means that
\beq
 \Bc'(1) = L \gamma_1 \Bc(1)  + O(\eps_L^2)
 \label{equ:p1}
\eeq
%where the approximate equality holds up to a correction proportional to $\eps_L^2$.  
Within the active phase, we ignore the exponentially small correction. For large $x$ the Airy function has
$\Bi'(x)/\Bi(x)\approx x^{1/2}$ from which one obtains
\beq
\mu^3+\psi_L \approx \gamma_1^2
\label{equ:mu3-psi-gamma}
\eeq
Recalling that  $F$ is small,  we have from (\ref{equ:bdy-Q-psi}) that $\gamma_1^2=F^2+O(F^3)$ which yields $\psi_L - F^2 \approx \mu^3 + O(F^3)$ or equivalently
\beq
\Psi_L \approx \Psi_{\rm a} = -\lambda\overline{k}
\eeq
which was the result anticipated in Fig.~\ref{fig:fss-sketch} for the active phase.  (The second equality is the definition of $\Psi_{\rm a}$, the first equality is approximate and holds up to corrections at $O(F^3)$.)
%Using (\ref{equ:mu-psi}) with (\ref{equ:gamma01}) to write $\gamma_1$ in terms of $\psi_L$ instead of $\Psi_L$, this formula reduces to a cubic  equation in $\psi_L$, where $L$ does not appear at all(?).   
%One of(?) the solutions to this equation is denoted by $\psi^{(1)}(\mu)$, and this gives the free energy of the active phase.
% and we expect that $\psi_L$ converges to $\psi^{(1)}(\mu)$ on taking $L\to\infty$ within the active phase.
%
%This equation can be written in terms of the Bi 
%function and solved for $\psi_L$ as a function of $\mu$.  
%The solution is denoted by $\psi^{(1)}_L(\mu)$.   It can be positive or negative.  
%It differs from $\psi_L$ by an exponentially small correction.
%
%
%In fact, for large $L$ we know that $\Bc(1)=1$ and we can use that for large $y$ then $\Bi'(y)/\Bi(y)\approx y^{1/2}$ to obtain
%\beq
%\mu (\mu+\psi_L/\mu^2)^{1/2} \approx \gamma_1
%\eeq
%This means that $\psi^{(1)}_L(\mu)$ converges for large $L$ to the solution of this equation.

%\emph{what can  we say about this phase?   We know that $\psi$ starts from $2(\cosh F-1)$ when $\lambda=0$ and decreases, we should presumably know that the the gradient wrt $\lambda$ is close to $1$.  At some point $\psi$ should go negative.}

\subsubsection{Inactive phase}
\label{subsec:p0}

For the inactive phase then $V_L$ is small which means that
\beq
L \gamma_0 \Ac(0) = \Ac'(0) + O(\eps_L^2)
 \label{equ:p0}
\eeq
As before we drop the term at order $\eps_L^2$ and obtain
\beq
L \gamma_0 \Ai( L^{2/3}\psi_L/\mu^2 )  = \mu L^{2/3} \Ai'( L^{2/3}\psi_L/\mu^2) 
\label{equ:Ai-L-gamma0}
\eeq
Recall that $\gamma_0 = F+O(F^2)$ is positive.  The derivative $\Ai'(x)$ is negative for $x>0$ so $\psi_L<0$. However, $\Ai'(x)$ oscillates for $x<\alpha_0$ where $\alpha_0\approx-2.3$ is the zero of $\Ai$ closest to $x=0$.   It follows that for large $L$, the argument of the Airy function must converge to $\alpha_0$ and one sees that
\beq
\psi_L \approx - |\alpha_0| \mu^2 L^{-2/3}
\label{equ:psiL-alpha0}
\eeq
Hence
\beq
\Psi_L(\lambda) \approx \Psi_{\rm i}(\lambda) = -2\sqrt{pq}F^2 + O(L^{-2/3})
\label{equ:Psi-i}
\eeq
as anticipated in Fig.~\ref{fig:fss-sketch}.  This behaviour was discussed in~\cite{Bodineau2012cmp,Bodineau2012jsp}.  The non-integer power in the finite-size correction reflects the fact that $\Ac_L(y)$ tends to zero as $y\to0$ but it has a maximum at a value of order $L^{-2/3}$, indicating that the typical value of $x=yL$ is large but subextensive, of order $L^{1/3}$.  That is, the inactive phase contains a single domain where the activity is non-zero: this domain covers a large number of sites, but not enough to constitute a finite fraction of the system~\cite{Bodineau2012cmp,Bodineau2012jsp}.

\subsubsection{Phase coexistence}

From the analysis of the two phases and recalling Fig.~\ref{fig:fss-sketch}, we see that the crossover between the phases takes place when $\Psi_{\rm a}\approx\Psi_{\rm i}$, which corresponds to $\mu=\mu_L^*$ with 
\beq
\mu_L^* \approx F^{2/3} 
\label{equ:mu*}
\eeq 
or equivalently
\beq
\lambda^*_L \overline k \approx 2\sqrt{pq} F^2
\label{equ:lambda*}
\eeq
consistent with Fig.~\ref{fig:fss-sketch}.  Note also  that $\psi_L^*\approx 0$ by (\ref{equ:psiL-alpha0}).
Within each individual phase, one of the factors $V_L$ and $W_L$ in (\ref{equ:VW}) is exponentially small, such that (\ref{equ:ab-1}) is satisfied.  Close to $\lambda^*$, both these quantities are small.  We analyse this situation in a general setting with minimal assumptions on the functions $V_L,W_L$.  The method is quite generic for first-order phase transitions: it amounts to constructing an eigenfunction $Q_L$ from two basis functions  (here $\Ac,\Bc$) which have exponentially small  overlap.  The aim is to compute the behaviour of the eigenvalue $\psi$ as the parameter $\mu$ is increased.

Consider (\ref{equ:p1}, \ref{equ:p0}), and drop the correction terms at $O(\eps_L^2)$.  These equations can be solved simultaneously for $(\mu,\psi)$.  We denote the solution by $(\mu_L^*,\psi_L^*)$, which corresponds to the point where $\Psi_{\rm a}=\Psi_{\rm i}$ in  Fig.~\ref{fig:fss-sketch}.  Note that if we substitute the values $(\mu_L^*,\psi_L^*)$ then $V_L=0=W_L$ exactly but (\ref{equ:VW}) is manifestly not satisfied in this case.  In fact, the true value of $\psi_L$ at this value of $\mu$ differs from $\psi_L^*$ by a correction at $O(\eps_L)$, recall again  Fig.~\ref{fig:fss-sketch}.

\begin{figure}
\begin{center}
\includegraphics[width=9cm]{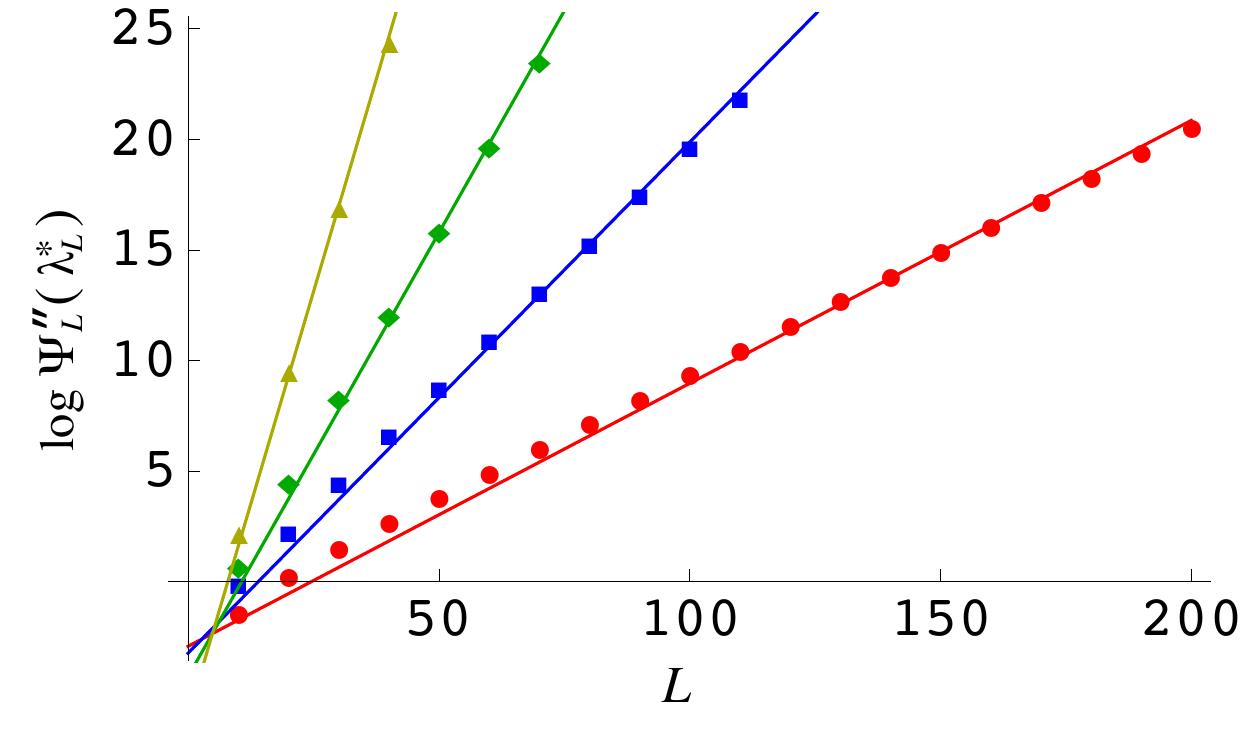}
\caption{The logarithm of the second derivative of SCGF at the transition point $\lambda_L^*$ as a function of the system size $L$: $\Psi_L''(\lambda^*_L)$. We numerically diagonalize the matrix of the largest eigenvalue problem (\ref{equ:mat-bulk}), evaluate the second derivative of the largest eigenvalue $\Psi_L(\lambda)$, and find the maximum of the second derivative $\Psi_L''(\lambda)$: $\max_{\lambda}\Psi_L''(\lambda)\equiv \Psi_L''(\lambda^*_L)$. Red circles, blue squared, green diamonds, yellow triangles correspond to $c=0.3,c=0.5,c=0.7,c=0.9$, respectively. We also plot the analytical prediction Eq.~(\ref{equ:chi*-predict}) as solid lines: $\log \Psi_L''(\lambda^*_L) = (2 F / 3) L + {\rm const.}$}
\label{fig:exponent}
\end{center}
\end{figure}

We are not able to estimate $(\mu_L^*,\psi_L^*)$ with exponential accuracy, but we are able to make an expansion about $\mu=\mu_L^*$ that captures the small deviation of $\psi_L$ from $\psi_L^*$. To this end, write
\beq
(\mu_L,\psi_L) = (\mu^*_L+\delta\mu,\psi^*_L+\delta\psi)
\eeq
and substitute in (\ref{equ:VW}). 
Using $\partial_\mu$ to indicate a partial derivative with respect to $\mu$, and similarly $\partial_\psi$, one obtains
\beq
\eps_L^2 \eta_L \xi_L = [ (\delta \mu) \partial_\mu W_L + (\delta \psi) \partial_\psi W_L ] \cdot [ (\delta \mu) \partial_\mu V_L + (\delta \psi) \partial_\psi V_L ]
\eeq
where all derivatives are evaluated at $(\mu^*,\psi^*)$.  It is not necessary to estimate these derivatives directly, instead we 
rewrite the equation in a generic form by introducing new constants $X_1,X_2,Z$ (dependent on the model parameters and on $\mu^*_L,\psi^*_L)$, such that
\beq
 ( \delta \psi + X_1 \delta \mu  ) \cdot (\delta \psi + X_2 \delta \mu ) -  \eps_L^2 Z^2 (X_2-X_1)^2  = 0
\eeq
Consistency of this approach requires $X_1\neq X_2$ and $Z^2>0$ but this is always the case in practice.  We take $X_2>X_1$ without loss of generality.
%Since $\eps_L$ is very small, this means that either $ \delta \psi \approx -X_L^{(0)} \delta \mu$ (which we identify  below as phase-0), or $\delta \psi \approx X_L^{(1)} -\delta \mu$ (which is phase-1), or both terms in brackets are small (which requires $\delta\mu$ to be of order $\eps_L$).
Solving for $\delta\psi$, we obtain
\beq
\delta\psi =  \pm \frac12 (Y-X) \sqrt{ (\delta\mu)^2 + (\eps_L Z)^2} -\frac{X+Y }{2} \delta \mu
\eeq
We take the $+$ sign for the square root because $\psi$ is known to be a convex function of $\mu$, see also below.  
Using (\ref{equ:mu-psi}) one has for small $\delta\mu$ that $\lambda-\lambda_L^* \approx C\delta\mu$ with $C=\frac{6\sqrt{pq}}{\overline{k} }(\mu^*_L)^2$.  Hence
\beq
\fl\quad
\frac{1}{2\sqrt{pq}} \Psi_L(\lambda) = \psi_L^* - F^2 + 
\frac{X_2-X_1}{C} \sqrt{ (\lambda-\lambda^*_L)^2 + (\eps_L Z C)^2} -\frac{X_1+X_2}{2C} (\lambda-\lambda^*_L)
\label{equ:Psi-coex}
\eeq
which is the finite-size scaling form for $\Psi$, valid for small $\lambda-\lambda_L^*$.
Differentiating with respect to $\lambda$ gives the behaviour of the order parameter near the transition:
\beq
{k}_L(\lambda) = 
-\Psi_L'(\lambda) = \frac{\sqrt{pq}}{C} \left[ (X_2+X_1) - \frac{(X_2-X_1)(\lambda-\lambda^*_L)}{\sqrt{ (\lambda-\lambda^*_L)^2 + (\eps_L Z C)^2}} \right].
\label{equ:psi'}
\eeq
This function is of the same form as (\ref{equ:k-first}), showing that the general analysis of Sec.~\ref{sec:privman} is consistent with the specific calculation of this section.
It shows a crossover at $\lambda=\lambda^*_L$, from a value $2\sqrt{pq}X_2/C$ (at small $\lambda$) to $2\sqrt{pq}X_1/C$ (at large $\lambda$).  Comparing with (\ref{equ:k-first}) one identifies these quantities as the activities of the two phases.
The width of the crossover is $1/(CZ\eps_L)$, which diverges exponentially with the system size when $L$ is large.  From (\ref{equ:k-first}) we identify this with $\ee^{\Gamma_sL}/(\Delta k)^2$ where $\Delta k$ is the difference in activity between the phases.  The maximal susceptibility is 
\beq
 \Psi_L''(\lambda^*_L) = \frac{X_2-X_1}{2ZC^2} \cdot \frac{1}{\eps_L}
\eeq
which indeed diverges exponentially with $L$, as anticipated in  Fig.~\ref{fig:fss-sketch}.   
The rate of this exponential divergence is fixed by the scaling of $\eps_L$.  Using (\ref{equ:mu*}) and that $\psi_L^*\approx 0$ one obtains from (\ref{equ:eps}) exponential divergence of the susceptibility:
\beq
 \Psi_L''(\lambda^*_L) \sim \ee^{\Gamma_{\rm s} L} , \qquad \Gamma_{\rm s} = 2F/3 + O(F^2)\; .
 \label{equ:chi*-predict}
\eeq
The exponential scaling is consistent with the general arguments of Sec.~\ref{sec:YggL} and we identify $\Gamma_{\rm s}$ as the surface tension.  
In Fig.~\ref{fig:exponent}, we compare (\ref{equ:chi*-predict}) with numerical results for $\Psi_L''(\lambda^*_L)$, showing good agreement. The analysis of this section also predicts the eigenfunction $Q$ which gives the probability distribution of $y_t$, at the coexistence point.  Note that the prediction (\ref{equ:chi*-predict}) for $\Gamma_{\rm s}$ is accurate up to corrections at $O(F^2)$. A more accurate numerical prediction is also available if one accounts for higher-order terms when solving (\ref{equ:mu3-psi-gamma}), see~\cite{Nemoto2017} for details.
%\emph{RLJ: For this work, I would prefer the simplicity of the result given here.} 

We conclude that the interfacial model predicts the scaling of the order parameter near the dynamical phase transition to be (\ref{equ:psi'}) and it also makes quantitative predictions including (\ref{equ:lambda*}, \ref{equ:chi*-predict}).  These have been shown~\cite{Nemoto2017,banuls2019} to agree very well with numerical computations on the FA model.  This confirms the prediction (\ref{equ:G-Psi}) that the interfacial model can capture the behaviour of the FA model close to its dynamical phase transition, including the relevant phase behaviour.

\section{Conclusion}
\label{sec:conc}

We have analysed the dynamical phase transition that occurs in the FA model, by two methods.  In Sec.~\ref{sec:numerics}, we invoked a correspondence between FA model trajectories and configurations of the $2d$ spin model, in order to relate dynamical properties of FA model to thermodynamic properties of this spin model.  We used the analysis of Privman and Fisher~\cite{Privman1983} to rationalise the finite-size scaling properties of this spin model, and we emphasised the differences in behaviour for systems with $Y\sim L$  and $Y\gg L$.  In the dynamical case, the first case corresponds to trajectories where the time $\tau$ and the system size $L$ are both large; the second case is relevant on taking the limit $\tau\to\infty$ before any limit of large $L$, as is often the case in studies of dynamical LDPs.

In this second case, we explained how the distribution of the order parameter at phase coexistence takes a unimodal form for large $\tau$, and the associated susceptibility diverges exponentially fast with $L$.  This is in contrast to the case $Y\sim L$ which is more conventionally analysed in thermodynamics, where the distribution is bimodal and the susceptibility diverges as a power law.
For $Y\gg L$, we also derived the general predictions (\ref{equ:LDP-gg}, \ref{equ:k-first}) for the finite-size scaling behaviour close to phase coexistence, based on the physical picture of Fig.~\ref{fig:privman-sketch}(b).
See also~\cite{Jack2019-colloq}.

In Section~\ref{sec:interfacial} we performed a specific analysis of the behaviour close to dynamical phase coexistence in  the FA model, using the interfacial model discussed in~\cite{Nemoto2017}, see also~\cite{Bodineau2012cmp,Bodineau2012jsp}.  The result is consistent with the general formulae (\ref{equ:LDP-gg}, \ref{equ:k-first}) but it also gives more detailed predictions, including the value of the surface tension parameter $\Gamma_{\rm s}$, and the forms of the eigenfunctions $Q$, which specify the probability distribution for the instantaneous size of the large inactive domain in the system.

The FA model is simple enough to enable this analysis, but the phase transitions that we have analysed exhibit the full phenomenology of first-order phase transitions in finite-dimensional systems.  We hope that these results will be useful in guiding the future analysis of other dynamical phase transitions and their finite-size scaling properties.

\ack 

VL is supported by the ERC Starting Grant No.~680275 MALIG, the ANR-18-CE30-0028-01 Grant LABS and the ANR-15-CE40-0020-03 Grant LSD.

\begin{appendix}

\section{MC simulation of the  $2d$ spin model}
\label{app:mc}

\newcommand{\CC}{{\cal C}}

To analyse thermodynamic properties of the $2d$ spin model, we use MC simulations.  Given a configuration ${\cal C}$ we propose a new configuration ${\cal C}'$.  The new configuration is accepted with probability $\min(1,{\rm e}^{E_s(\CC)-E_s(\CC')})$ where the energy $E_s$ is given by (\ref{equ:Es}), generalised to periodic boundaries if applicable.  Otherwise the old configuration is retained.

In the simplest MC algorithm, one chooses either a spin $n_{i,y}$ or $m_{i,y}$ at random and one proposes a new configuration by flipping this spin.    
However, for systems with periodic boundaries, the constraints on the $n$ and $m$ variables means that updating the configuration in this way does not allow access to every possible configuration of the system.  As an example, consider the following configuration of a periodic system with $(L,Y)=(4,5)$, where $\tt X$ and $\tt .$ indicate sites with $n_i=1,0$ respectively and $\tt-$ indicates a bond with $m_i=1$.  
\begin{verbatim}
 X X X X X
 X-. .-X X
 X X-. .-X
 X X X X X
\end{verbatim}

It can be checked that there is no sequence of single-spin flips that connect this configuration to the state where $n_{i,y}=1$ for all $i,y$.  The essential problem is that there are two spins in the third column with $n_{i,y}=0$ (indicated by $\tt .$), but  neither of these spins is able to flip, because of the constraint that $m_{i\pm1}=0$ if $m_{i}=1$.

The solution is that in addition to proposed moves where only one spin is flipped, we also propose occasionally moves where two spins $n_{i,y}$ and $n_{i,y+1}$ are both flipped simultaneously.  We claim that this gives an MC method that samples the full configuration space of the model.  For small  systems we have verified explicitly every allowed configuration can be reached from every other configuration. 

With this MC algorithm, the system explores its configuration space slowly, but computations are feasible.  For  calculations in large systems close to phase transitions (Figs.~\ref{fig:L32}, \ref{fig:L16}), we use two methods.  For $L=16$ we performed long simulations with $s\approx s^*$, so that the system visits both phases many times during the coarse of a single MC run.  A histogram of $K$ is accumulated in this run, and $k(s)$ and $\chi(s)$ are obtained as weighted average with respect to that histogram.  For $L=32$ we performed parallel tempering simulations in which multiple replicas of the system are simulated at different values of $s$, with all values of $s$ being close to $s^*$.  The histograms from the data at different $s$ are combined using the unbinned weighted histogram analysis method~\cite{Tan2012}, which yields an estimate for the distribution of $K$.  This distribution is then used to derive $k(s)$ and $\chi(s)$ by weighted averages.

\begin{figure}
\includegraphics[width=150mm]{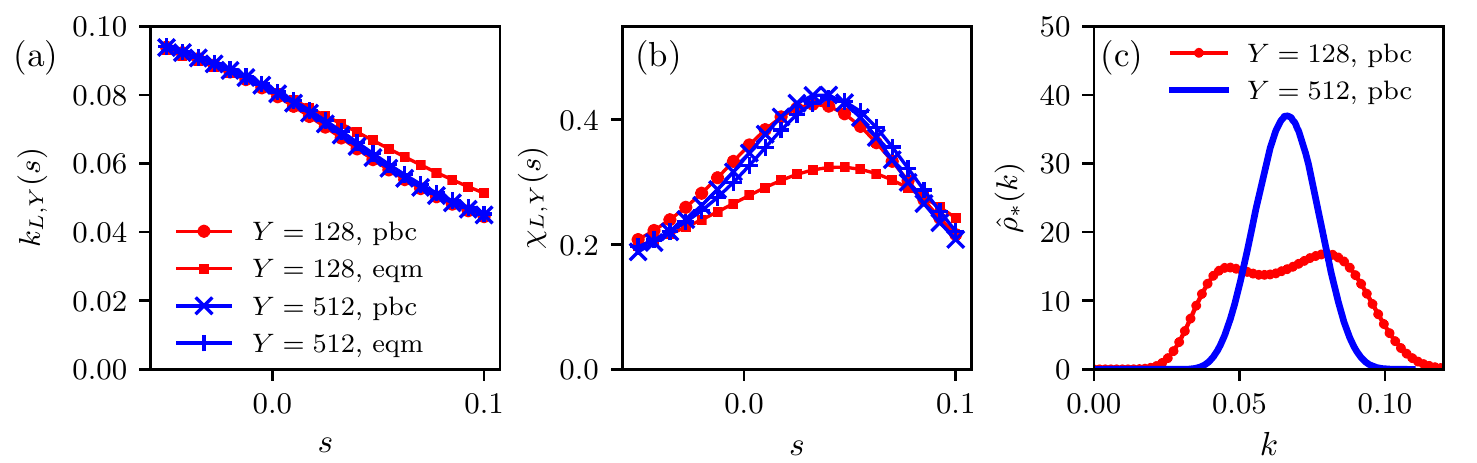}
\caption{Data for the $2d$ spin model at $L=8$.  We show results for systems with periodic boundaries (pbc), and for systems with energy (\ref{equ:Es}), which includes boundary terms that are chosen to recover the equilibrium trajectory ensemble of the dFA model (eqm).  (a) Order parameter $k_{L,Y}(s)$. (b) Susceptibility $\chi_{L,Y}(s)$, the colours and symbols are the same as in panel~(a). (c) Finite size scaling of $\hat\rho_*$, only for the periodic case.  Even if the function $k_{L,Y}(s)$ has saturated to its large-$Y$ limit, the shape of this distribution still depends on $Y$.}
\label{fig:L8}
\end{figure}

Finally, to demonstrate the effect of different boundary conditions, Fig.~\ref{fig:L8} shows data for a very small system ($L=8$).  Taking $Y\to\infty$ at fixed $L$, one expects convergence of $k(s)$ and  $\chi(s)$ to smooth limiting forms at large  $Y$.  The figure shows that $Y=128$ is already very close to convergence of this limit in a system with periodic boundaries.  By contrast, the system which corresponds exactly to the equilibrium trajectory ensemble of the dFA model has a much stronger finite-size effect, in that larger $Y$ is required to see convergence to the large-$Y$ limit.  The reason is that the regions close to $y=1$ and $y=Y$ are biased by the boundary conditions towards the active phase, which hinders characterisation of the phase coexistence regime.  See also~\cite{Elmatad2010}.

\section{Simple model for multi-domain phase coexistence}
\label{app:1d}

We describe the fluctuations in configurations such as those sketched in Fig.~\ref{fig:privman-sketch}(b).  These are representative configurations of the $2d$ spin model at $s=s^*$.  They  can be modelled by considering a $1d$ Ising model with periodic boundaries whose typical domain size is $\kappa \approx \ee^{\Gamma_{\rm s}L}\gg1$.  The same results may alternatively be obtained by considering large deviations of a two-state Markov chain.  Working at temperature $T=1$ we introduce a magnetic field $h\ll 1$ so the free energy of the Ising model can be obtained by standard methods as $\lambda(h)=-\frac1Y\log\mathrm{Tr}(M_h^Y)$ where the transfer matrix is
\beq
M_h =  \left( \begin{array}{cc}  1+h& \kappa^{-1} \\  \kappa^{-1} &  1-h \end{array} \right) \; .
\eeq
where we work at leading order in $h,\kappa^{-1}$.  
For large $Y$ then $\lambda(h)$ is given by (the negative of) the largest eigenvalue of this matrix so $\lambda(h)=-1-\sqrt{h^2+\kappa^{-2}}$. The SCGF for the total magnetisation of the $1d$ Ising model (in this limit) is $g(h)=\lambda(0)-\lambda(h)$  so
\beq
g(h) = \sqrt{h^2+\kappa^{-2}}  - \kappa^{-1}  \; .
\eeq  
The derivative of this function gives the mean magnetisation as a function of the field
\beq
m(h) = g'(h) = \frac{h}{\sqrt{h^2+\kappa^{-2}}} \; .
\label{equ:mh-1d}
\eeq
The total magnetisation obeys an LDP whose speed is the number of sites in this $1d$ Ising model, which is $Y$.
The rate function for this total magnetisation is ${J}(m) = \sup_h( hm - g(h))$ which yields
\beq
J(m) = \kappa^{-1} ( 1 - \sqrt{1-m^2} ) \; , \qquad -1 < m< 1\; .
\label{equ:rate-J}
\eeq
This rate function describes the fluctuations of the amount of each phase, assuming that the domain walls in the configuration of Fig.~\ref{fig:privman-sketch}(b) are a dilute ideal gas.  
Neglecting fluctuations within the phases, we identify the (intensive) activity of a configuration as $k = k_0 + (m\Delta k/2)$ where $k_0$ is the average activity of the two phases, and $\Delta k$ is the difference between their activities.  Hence (\ref{equ:rate-J}) yields (\ref{equ:LDP-gg}, \ref{equ:triv-ldp}).  

%Note also, the Boltzmann factor that appears in the $1d$ Ising model is $h\sum_y m_y$ whose analogue in the $2d$ spin  model is $sL\sum_{y}k_y$, where $k_y$ is the fraction of spins that change their state between states $\bm{n}_y$ and $\bm{n}_{y+1}$.  Identifying 
%
%Similarly one obtains (\ref{equ:k-first}) from \ref{equ:mh-1d} by identifying $h$ with $-(s-s^*)\Delta k/2$.  The offset $s^*$ enters because the $1d$ Ising calculation describes fluctuations about the symmetric state illustrated in (\ref{fig:privman-sketch}(b), which is characteristic of the  phase coexistence point $s=s^*$.

\section{Discrete and continuous versions of the interfacial model}
\label{app:continuum}

\subsection{Continuum limit}

\begin{figure}
\includegraphics[width=150mm]{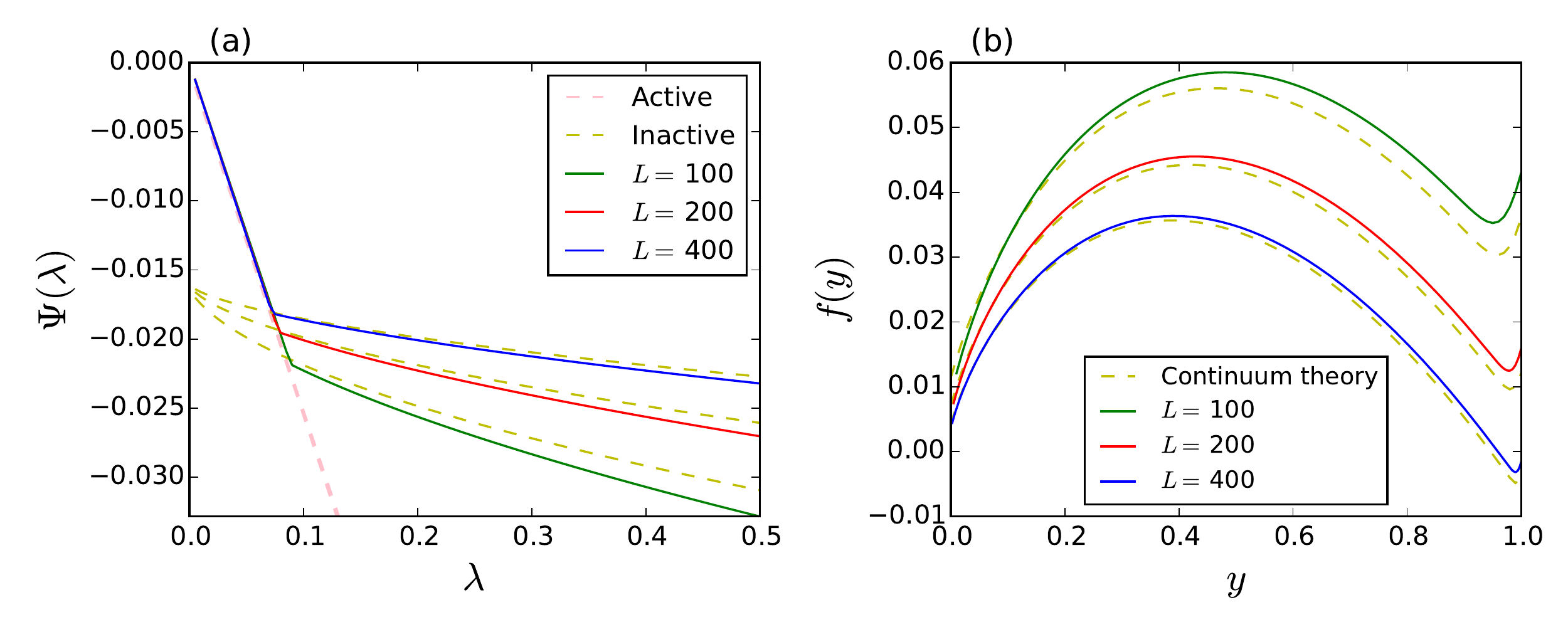}
\caption{{\bf (a)} The SCGF $\Psi(\lambda)$ obtained 
from the discrete eigenvalue problem  (\ref{equ:mat-bulk}) (solid lines) and 
continuous eigenvalue problem (\ref{equ:airy-Q}) (dashed lines). The value of $c$ is 0.3. The active and inactive solutions for the continuous problem are obtained following the procedure detailed in the main text. {\bf (b)} The behaviour of $f(y) = (1/L) \log P_L(y)$ in the inactive phase, for $(c,\lambda)=(0.3,0.2)$.  We show good agreement between solutions of the  discrete problem [Eq.~(\ref{equ:mat-bulk}), solid lines] and those obtained by solving the continuous problem [Eq.~(\ref{equ:airy-Q}), dashed lines]. 
%The values of $c$ and $\lambda$ are set to $0.3$ and $0.2$, respectively.
}
\label{fig:continuous_exact}
\end{figure}

We show how the interfacial model (defined on a discrete lattice) can be analysed by considering a continuous probability distribution for $y$.
Starting from (\ref{equ:mat-bulk}) and using (\ref{equ:PLf}) we obtain (after Taylor expansion of $f$ with $L\gg 1$):
\beq
\fl \quad \Psi +\lambda y \bar k = 2p \left( \ee^{f'(y)}[1+f''(y)/(2L)] - 1 \right) + 2q  \left( \ee^{-f'(y)}[1+f''(y)/(2L)] - 1 \right). 
\eeq
So far this result only requires that $L$ is large.  In order to obtain an eigenvalue problem for the continuous function $P$, we now assume additionally that $|f'(y)|\ll 1$ everywhere, which leads to
\beq
\Psi  +\lambda y \bar k = (2p-2q) f'(y) + (p+q) [ f'(y)^2 + f''(y)/L ].
\eeq
We will see that  this corresponds physically to approximating the binomial distribution of the discrete random walk  by a Gaussian distribution, as in a diffusion process.  When considering the typical behaviour of the system, this approximation is exact for large $L$, but large deviations can be sensitive to the full distribution of hop sizes.  For the problem considered here, we make an approximation when expanding over $f'$ but the resulting theory still gives semi-quantitative predictions (see Fig.~\ref{fig:exponent} and Fig.~\ref{fig:continuous_exact}).
Identifying 
\beq
Lf'(y)=\frac{\mathrm{d}}{\mathrm{d}y}\log P(y) = \frac{P'(y)}{P(y)}
\label{equ:Lf'}
\eeq
and
\beq
Lf''(y)= \frac{\mathrm{d}^2}{\mathrm{d}y^2}\log P(y)  = \frac{P''(y)}{P(y)} - \frac{P'(y)^2}{P(y)^2}, 
\eeq one obtains
\beq
(\Psi + \lambda y)P(y) = \frac{2(p-q)}{L} P'(y) + \frac{p+q}{L^2} P''(y)
\label{equ:P-eigen-app}
\eeq
which is an eigenvalue equation for the continuous function  $P$.  
The boundary conditions (\ref{equ:mat-bdy}) are treated similarly, in order to obtain constraints on $f'(0)$ and $f'(1)$.  However, the point about which the Taylor expansion is performed is optimised to minimise errors associated with the expansion at small $f'$.  To this end note that
\beq
\fl\quad \frac{P((n+1)/L)-P(n/L)}{P((n+1)/L)+P(n/L)} = \frac{\ee^{f'}-1}{\ee^{f'}+1} = \tanh(f'/2) = \frac{f'}{2} + O(f')^3
\label{equ:tanh-f}
\eeq
where we use the shorthand $f'=f'(\frac{2n+1}{2L})$, for compactness of notation.  
The correction on the right hand side is $O(f')^3$: other representations of the derivative are possible but would have corrections at $O(f')^2$ which is less accurate.
The boundary conditions (\ref{equ:mat-bdy}) are 
\beq
\frac{P(2/L)-P(1/L)}{P(2/L)+P(1/L)} = c_0 , \qquad \frac{P(1)-P((L-1)/L))}{P(1)+P((L-1)/L))} = c_1
\eeq
with $c_0=\frac{2q-2p+\Psi_L}{2q+2p+\Psi_L}$ and $c_1=\frac{2q-2p-\Psi_L-\lambda\overline{k}}{2q+2p+\Psi_L+\lambda\overline{k}}$.  Hence (\ref{equ:tanh-f}) provides values for $f'(\frac{3}{2L})$ and $f'(\frac{2L-1}{2L})$ which we identify (at large $L$) with $f'(0)$ and $f'(1)$.  Using again (\ref{equ:Lf'})  yields 
\beq
\frac{P'(0)}{P(0)} = 2c_0 L, \qquad \frac{P'(1)}{P(1)} = 2c_1 L \; .
\label{equ:P-bdy}
\eeq

\subsection{Simplification for small $F$}

Note that in addition to a large-$L$ limit, we have assumed that $f'(y)\ll1$.  For $\lambda=0$ it is easily verified that the exact eigenvector of the discrete problem has $f'(y)=2F=\log(q/p)$ so  self-consistency requires that this parameter be small.  Physically, this requires that the hop rate for the asymmetric random walk $x_t$ is much larger than its drift,
that is $(p-q)/(p+q)\ll 1$.

In this case it is consistent to identify $(p-q)=2\sqrt{pq}\sinh F\approx 2F\sqrt{pq}$ and $(p+q)=2\sqrt{pq}\cosh F\approx (2+F^2)\sqrt{pq}$.  Using this result in (\ref{equ:P-eigen-app}) yields (at leading order)
\beq
\frac{1}{2\sqrt{pq}} (\Psi + \lambda \overline{k} y)P(y) = \frac{2F}{L} P'(y) + \frac{1}{L^2} P''(y)
\label{equ:P-eigen-F}
\eeq
and the associated boundary conditions are
\beq
\fl\qquad
\frac{P'_L(0)}{P_L(0)} = 2L\cdot \frac{F\sqrt{pq}+(\Psi_L/4)}{\sqrt{pq}+(\Psi_L/4)} , \qquad 
\frac{P'_L(1)}{P_L(1)} = 2L \cdot \frac{F\sqrt{pq}-(\Psi_L+\lambda\overline{k})/4 }{\sqrt{pq}+(\Psi_L+\lambda\overline{k})/4}.
\label{eq:boundary_appendix}
\eeq
To see that the small-$F$ approximation is self-consistent, note that for $\lambda=0$ one has $\Psi_L=0$ and this recovers $P(y)=\ee^{2LFy}$, which is the exact result in the discrete case.  

To transform the eigenvalue problem to a self-adjoint form we define $Q(y)=P(y)\ee^{-LFy}$, which yields
\beq
\frac{1}{2\sqrt{pq}} (\Psi + \lambda \overline{k} y)Q(y) = \frac{1}{L^2} Q''(y) - F^2 Q(y) \; .
\label{equ:Q-eigen-F}
\eeq
As $L\to\infty$, one sees (consistent with Fig.~\ref{fig:fss-sketch}) that both $\Psi$ and $\lambda \overline{k}$ will be $O(F^2)$.  Hence the boundary conditions for $Q$ can be derived as
\beq 
\fl
\frac{Q'(0)}{Q(0)} = L \left( F+\frac{\Psi}{2\sqrt{pq}} + O(F^3) \right) , \qquad \frac{Q'(1)}{Q(1)} = L \left( F-\frac{\Psi+\lambda\overline{k}}{2\sqrt{pq}} + O(F^3) \right) .
\label{equ:bdy-Q-F}
\eeq

\subsection{Numerical demonstrations}
\label{Appendix:numerical_demonstrations}

We use numerical results to compare solutions of the continuous eigenvalue problem (\ref{equ:airy-Q}) and the original discrete eigenvalue problem (\ref{equ:mat-bulk}), in order to demonstrate the validity of the continuous limit.

For the discrete problem, we numerically diagonalize the matrix on the right-hand side of (\ref{equ:mat-bulk}) and obtain $\Psi(\lambda)$ as the largest eigenvalue and $P_{L}(y)$ as the associated eigenvector. For the continuous problem, we use the general solution (\ref{equ:QL}) of the eigenvalue problem (\ref{equ:airy-Q}, \ref{equ:bdy-Q-psi}), where the parameter $\hat b_{L}$ is set to 1 without loss of generality and the parameter $\hat a_L$ is determined using one of the boundary conditions (\ref{equ:airy-Q}) and (\ref{equ:bdy-Q-psi}). The other boundary condition is used to determine the value of $\psi$ (as the largest solution that satisfies the boundary condition), from which we obtain $\Psi(\lambda)$. 
There are two technical remarks related to the boundary conditions. First, depending on which boundary condition is used to determine $\hat a_L$, the obtained solution $\psi$ changes: if the boundary condition for $y=0$ (or $y=1$) is used to determine $\hat a_L$, we obtain the solution of $\psi$ that corresponds to the active (or inactive) phase. This means that, irrespective of the value of $\lambda$, we can construct both active and inactive solutions by exploiting this property (see Fig.~\ref{fig:continuous_exact}, the solution with larger $\psi$ gives the free energy and the other solution corresponds to a metastable state).  Second, if we ignore the $O(F^3)$ terms in the boundary conditions (\ref{equ:airy-Q}) and (\ref{equ:bdy-Q-psi}), we may obtain an artificial solution in some cases (with $\psi>0$). To avoid this problem, we use the full expression of the boundary conditions ({\it i.e.}, the expression that includes $O(F^3)$ terms\footnote{The complete boundary condition that includes $O(F^3)$ can be derived from (\ref{eq:boundary_appendix}), it is also provided in Supplementary Material of Ref.~\cite{Nemoto2017}.}). Alternatively, this problem can be avoided by retaining the truncated boundary conditions and restricting to solutions with $\psi<0$. 

Fig.~\ref{fig:continuous_exact}(a) shows $\Psi(\lambda)$ for these discrete and continuous problems.  Fig.~\ref{fig:continuous_exact}(b) shows $f(y)=(1/L)\log P_{L}(y)$ for a representative state point in the inactive phase.  For large-$L$, there is good agreement between the solutions of the discrete and continuous eigenvalue problems, as expected.   Recall that we derived the continuous eigenvalue equation (\ref{equ:airy-Q}) from the discrete eigenvalue equation (\ref{equ:mat-bulk}) by assuming that the slope of $f(y)$ is small.  Our numerical observation shows that $f(y)$ has a zero slope for $y\sim 0.5$, and keeps a relatively small slope around $y\sim 0.5$ (except for around $y=0$ or $y=1$), which
 is consistent with this small-$f'(y)$ assumption. 
 Fig.~\ref{fig:continuous_exact} shows results for $c=0.3$; qualitatively similar results were also obtained in the range of $c$ between $c=0.1$ and $c=0.7$ (not shown).
 
To understand the physical interpretation of the results in Fig.~\ref{fig:continuous_exact}(b), recall from Sec.~\ref{subsec:eval-cts} that the maximum of $f$ corresponds to the most likely value of $y_\tau$, where $\tau$ is the final time of the trajectory.  The different behaviours of $P_L$ and $Q_L$ [defined in Eqs.~(\ref{equ:PLf}, \ref{equ:QLf})] mean that the most likely value of $y_\tau$ differs from the typical values of $y_t$ in the bulk of the trajectory (far from initial and final times).  The average values of $y_t$ in the bulk of these trajectories are (0.117, 0.0771, 0.0506) for $L=(100,200,400)$; these small values are consistent with the system being in the inactive phase.
%Note that this value is significantly larger than typical values of $y_t$ for times $t$ in the bulk of the trajectory, because of the differences between $P_L$ and $Q_L$ in (\ref{equ:PLf},\ref{equ:QLf}).  

\end{appendix}

\section*{Bibliography}
\bibliographystyle{plain_url}
\bibliography{dev}

\end{document}